# Flexible Wearable Filtering Antenna With Stable Performance for IoT Devices

Runkai Song, *Student Member, IEEE*, Fan Qin, *Member, IEEE*,
Wenchi Cheng, *Senior Member, IEEE*, and Steven Gao, *Fellow, IEEE*

***Abstract*—In this article, a flexible and lightweight filtering wearable antenna without extra circuits is presented. The proposed antenna starts from a flexible directional antenna with lightweight structure, which includes a layer of ultra-thin flexible substrate, a metal ground layer, and a flexible foam layer sandwiched between them. Then, by introducing two pairs of vertical slots to the radiation patch printed on the flexible substrate, two radiation nulls are realized at both band edges without extra circuits. Moreover, to mitigate the deterioration of in-band radiation under different curvature, a pair of inverted F-shaped slots are loaded on the radiation patch. The coupling of F-shaped slots suppresses non-radiated lateral current components along the curvature direction, maintaining stable performance after bending. In addition, deformation analysis of the proposed antenna with a three-layer human tissue model under different bending radii is carefully carried out, showing stable bandwidth, effective out-of-band radiation suppression, and low specific absorption rate (SAR) value. To verify this method, a prototype is fabricated. Measurements are conducted both in free space and conformal on the curved body tissue. The results show that the proposed antenna achieves a −10 dB impedance bandwidth from 2.7 GHz to 3 GHz, an out-of-band radiation suppression more than 11 dB with maxmium suppression of 23 dB, and an average gain of 8.5 dBi. As a flexible wearable antenna with stable performance and integrated reliable filtering features, it has several advantages including flexible wearable structure, stable filtering properties, lightweight characteristic, and low SAR. This makes it an excellent candidate for wearable IoT applications.**

## I. INTRODUCTION

IN recent years, with high integration, wearability, and compact size, wearable devices are being widely utilized in Internet of Things (IoT) scenarios [1]-[2]. Serving as skin-attached devices that continuously and closely monitor a person's behavior without interfering with or limiting the user's movements [3], the development of wearable IoT devices facilitates several innovative applications, such as equivalent daylight illuminance monitoring [4], health monitoring [5], real-time wearable fall detection system [6],

Manuscript received 26 October 2025; revised 1 December 2025; accepted 4 February 2026. This work was supported by the National Key R&D Program of China under Grant 2024YFC3016000. *(Corresponding author: Fan Qin.)*

Runkai Song, Fan Qin, and Wenchi Cheng are with the School of Telecommunications Engineering, Xidian University, Xi'an 710071, China (e-mail: srk@stu.xidian.edu.cn; fqin@xidian.edu.cn; wccheng@xidian.edu.cn).

Steven Gao is with the Department of Electronic Engineering, Chinese University of Hong Kong, Hong Kong (e-mail: scgao@ee.cuhk.edu.hk).

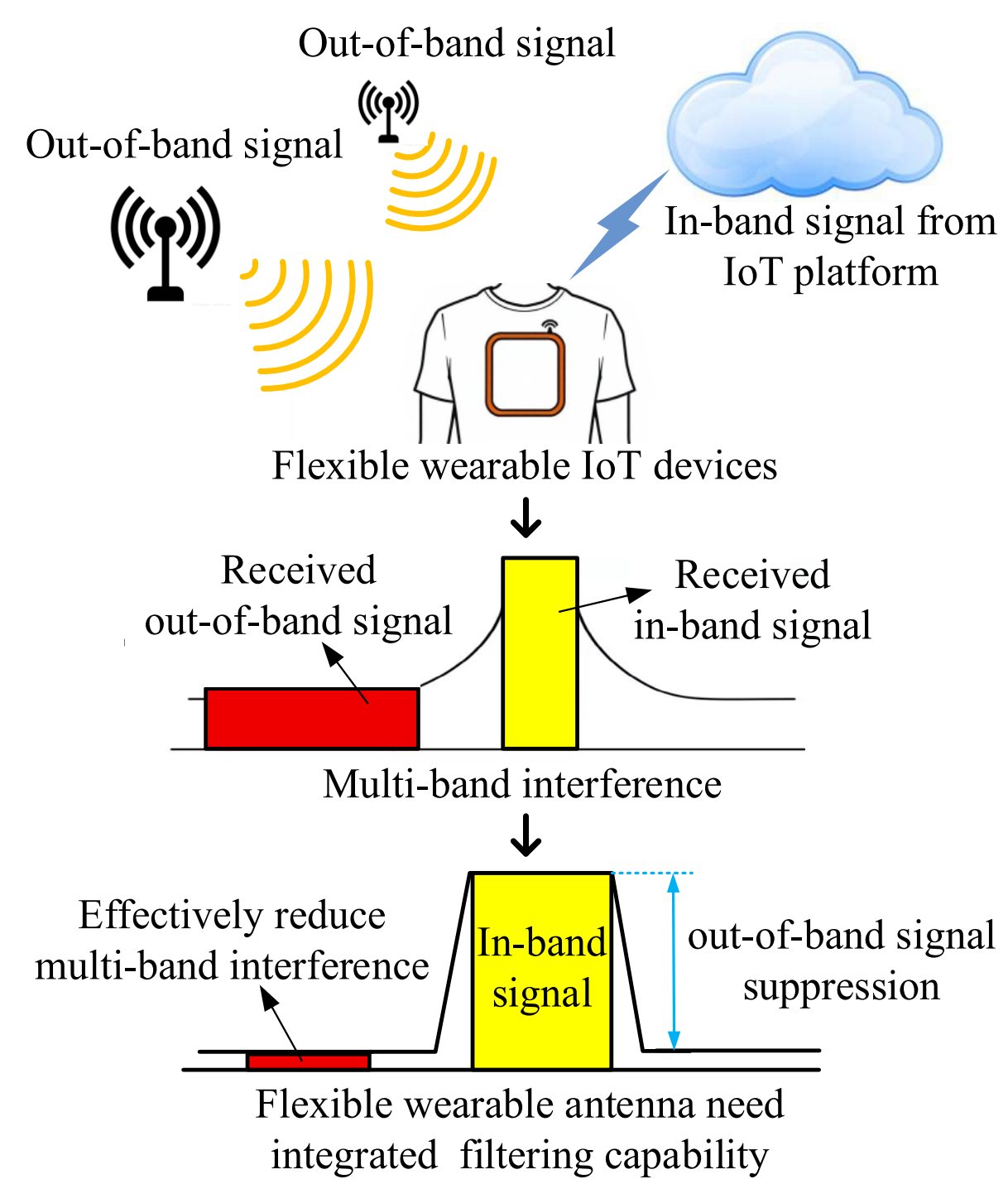

**Fig. 1.** Schematic of wearable devices transmitting diverse real-time data to the cloud on the IoT platform.

medical devices [7], human activity recognition [8], and digital forensics [9]. With continuous advancements in wearable IoT technology, an increasing number of wearable devices are utilized [10], becoming a critical component in improving IoT development.

Furthermore, as shown in Fig. 1, a large number of real-time data are transmitted from IoT platform to wearable devices. However, different types of wireless devices in the scenario are using various frequency bands to support their specific function as well. Multi-band interference generated by diverse out-of-band signals will seriously affect the stability of wireless communications between devices and IoT platform. Therefore, integrated out-of-band signal suppression ability of wearable IoT antennas is essential for reliable connectivity. Besides, for industrial private networks demanding low latency, stringent safety and high speed, the frequency band from 2.7 GHz to 2.9 GHz is being utilized to avoid interference from traditional frequency band. Moreover, wearable devices are required to adopt flexible structure and stable on-body performance. Furthermore, lightweight design for comfortable wearing experience is also essential. To better address the evolving requirements of wearable devices, wearable antennas integrating filtering capabilities, flexibility, and lightweight characteristic are highly desirable.

In the past, the combination of thin profile, compact size, conformal capability, and flexibility has spurred the development of numerous wearable antenna designs [11]-[13]. In [14], a dynamic decoupling network designed for wearable antennas is reported. A miniaturized wearable annular slot antenna introduced in [15] is achieved by loading rectangular corrugated metal-insulator-metal structures. These antennas achieve compact size and low specific absorption rate (SAR) value. However, the absence of flexible design severely limits their wearing experience. In [16], a flexible wearable circularly polarized antenna which consists of a feed branch, coupling branch, and peripheral annular ground is introduced. In [17], a pattern-reconfigurable wearable antenna is achieved by attaching a rectangular slot to the radiation patch and mounting four pin diodes. With flexible structure, these wearable antennas can maintain stable on-body performance after bending. However, they do not have filtering capability.

Moreover, with low insertion loss, compact size and multifunctional capabilities, research on filtering antennas has received growing attention [18]-[22]. Several approaches have been reported, such as inserting extra filtering circuits into feeding network [23]-[25], and integrating filtering design at the radiator [26]-[30]. In [24], the filtering performance is achieved by employing a dual-mode stub loaded resonator to the feeding networks. The filtering design reported in [25] introduces U-slot patch as a dual-mode resonator for filtering capability. Due to the applications of extra filtering circuits, these antennas introduce extra insertion loss and degradation of the in-band gain. To solve this problem, antenna designs with integrated filtering structure are reported. The filtering design introduced in [26] is realized by exciting electric and magnetic coupling on a folded dipole. Recently, a flexible dual-band filtering antenna is reported in [30]. This antenna achieves integrated filtering design with flexible structure for the first time. Additionally, some filtering wearable antenna designs have been reported in [31] and [32]. These antennas achieve low SAR and effective filtering capability.

However, to the best of our knowledge, none of the existing filtering wearable antennas have flexibility, which severely limits their wearing comfort and application scenarios. Besides, extra filtering circuits are applied for filtering functionality. Moreover, the existing flexible filtering antennas are not suitable for wearable applications because they cannot achieve low SAR and stable on-body bending performance. Consequently, it is necessary to design a flexible wearable filtering antenna without extra circuits for further development of wearable IoT technologies.

In this article, we propose a design of flexible filtering wearable antenna with stable on-body performance and low SAR. Filtering capability is achieved by introducing two pairs of vertical slots to the radiation patch. The design technique of stable on-body performance after bending is explained by loading a pair of inverted F-shaped slots on the radiation patch to suppress non-radiated lateral current components along the curvature direction. The flexible lightweight characteristic and low SAR value are realized through a double-layer flexible structure with a whole metal ground. With stable and effective filtering performance, flexibility, lightweight characteristic and low SAR value, the proposed wearable antenna can significantly mitigate multi-band interference between various wireless wearable IoT devices.

This paper presents three main innovations. Firstly, the proposed flexible wearable filtering antenna achieves stable on-body performance under different curvature by suppressing lateral current components along the curvature direction. Besides, the proposed design realizes integrated filtering design on a wearable antenna structure without extra circuits. Moreover, lightweight and low SAR designs are considered into antenna structure for wearable applications.

## II. Antenna Design And Simulation

### *A. Design Procedure and Antenna Configurations*

The structure evolution and prototype of the proposed antenna design are shown in Fig. 2. The design procedures from Ant. 1 to the proposed antenna are as follows.

1) Ant. 1: Baseline structure. The antenna has directional radiation with metal ground and a circular radiation patch.

2) Ant. 2: Radiation null at the lower band edge. A radiation null is achieved with attaching a pair of longer slots whose resonant frequency locates at the lower band edge. The main E-field with the reverse phase will be perpendicular to slots at the frequency of nulls, leading to low radiation efficiency.

3) Ant. 3: Filtering capability without bending. By adding a pair of shorter slots on the outside of longer slots, the other radiation null is produced at higher band edge. Then, Ant. 3 achieves effective out-of-band radiation suppression with two radiation nulls. However, its performance degrades after bending significantly. The in-band reflection coefficient of Ant. 3 deteriorates with increasing curvature, along with realized gain reduces correspondingly. Therefore, how to achieve stable in-band radiation after bending is essential.

4) The proposed antenna: Stable performance after bending. A pair of inverted F-shaped slots are loaded onto the radiation patch. The coupling of F-shaped slots effectively suppresses non-radiated lateral current components along the curvature direction, achieving stable and effective in-band radiation. The -10 dB impedance bandwidth of the proposed antenna maintains stable under different curved radii. Besides, the proposed antenna exhibits reliable in-band gain and filtering capability when conforming to the curved body.

The prototype of the proposed filtering antenna is shown in Fig. 2. The antenna structure consists of the upper radiation patch, a flexible polyimide (PI) flexible substrate, a flexible foam layer and the metal ground behind. The radiation patch is printed on the front PI flexible material with a thickness of 0.1 mm. The flexible foam layer with a thickness of 6.5 mm connects metal ground and the PI layer. The applications of PI material and foam substrate significantly improve the antenna's lightweight characteristic and conformal capability, which is essential in the wearable applications.

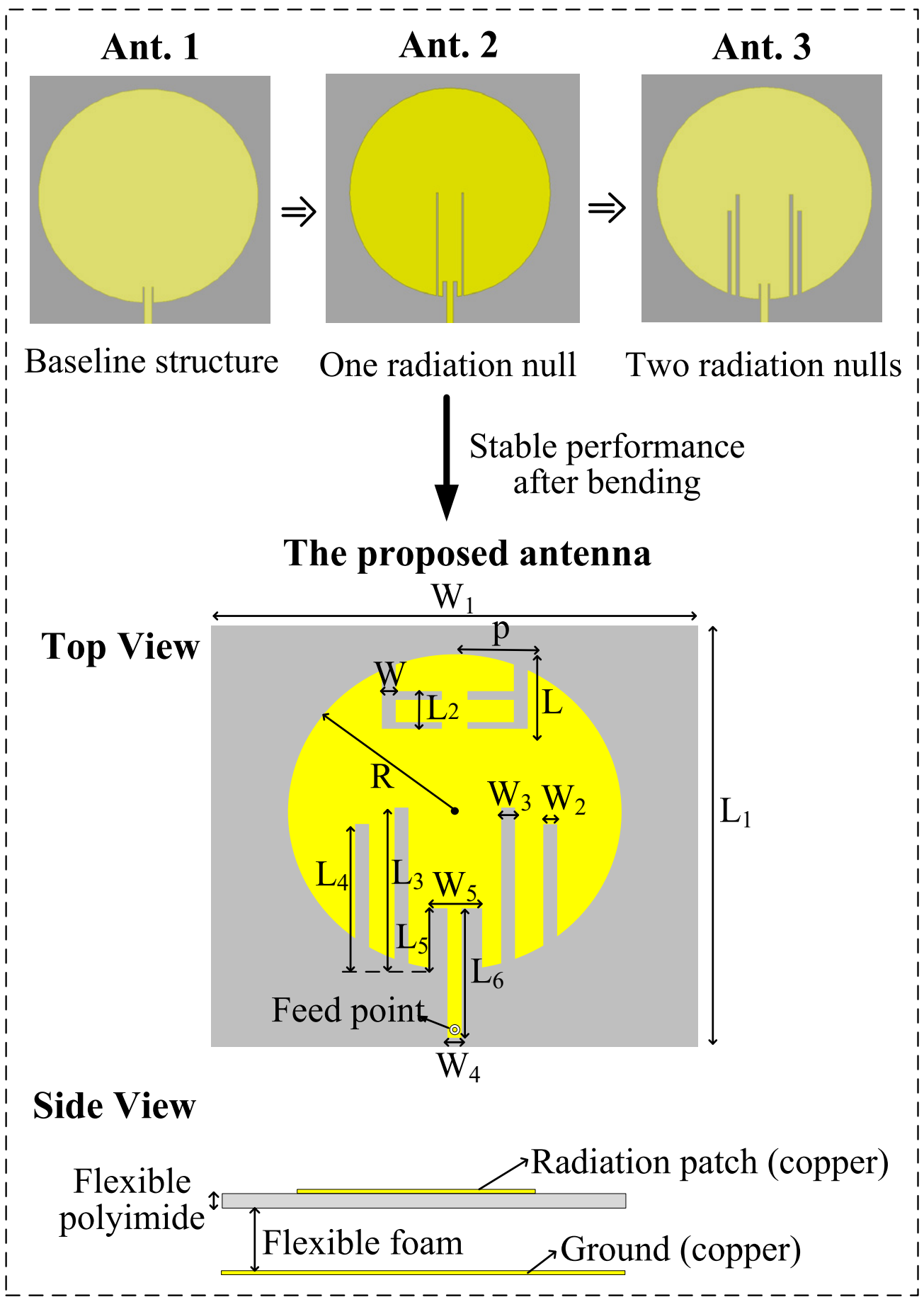

**Fig. 2.** Design procedure and prototype of the proposed antenna (with L = 10 mm, $L_1$ = 76.2 mm, $L_2$ = 6.8 mm, $L_3$ = 26 mm, $L_4$ = 23 mm, $L_5$ = 4 mm, $L_6$ = 20 mm, W = 1 mm, $W_1$ = 85 mm, $W_2$ = 1.1 mm, $W_3$ = 1.3 mm, $W_4$ = 1.7 mm, $W_5$ = 2.7 mm, R = 26 mm, p = 10 mm).

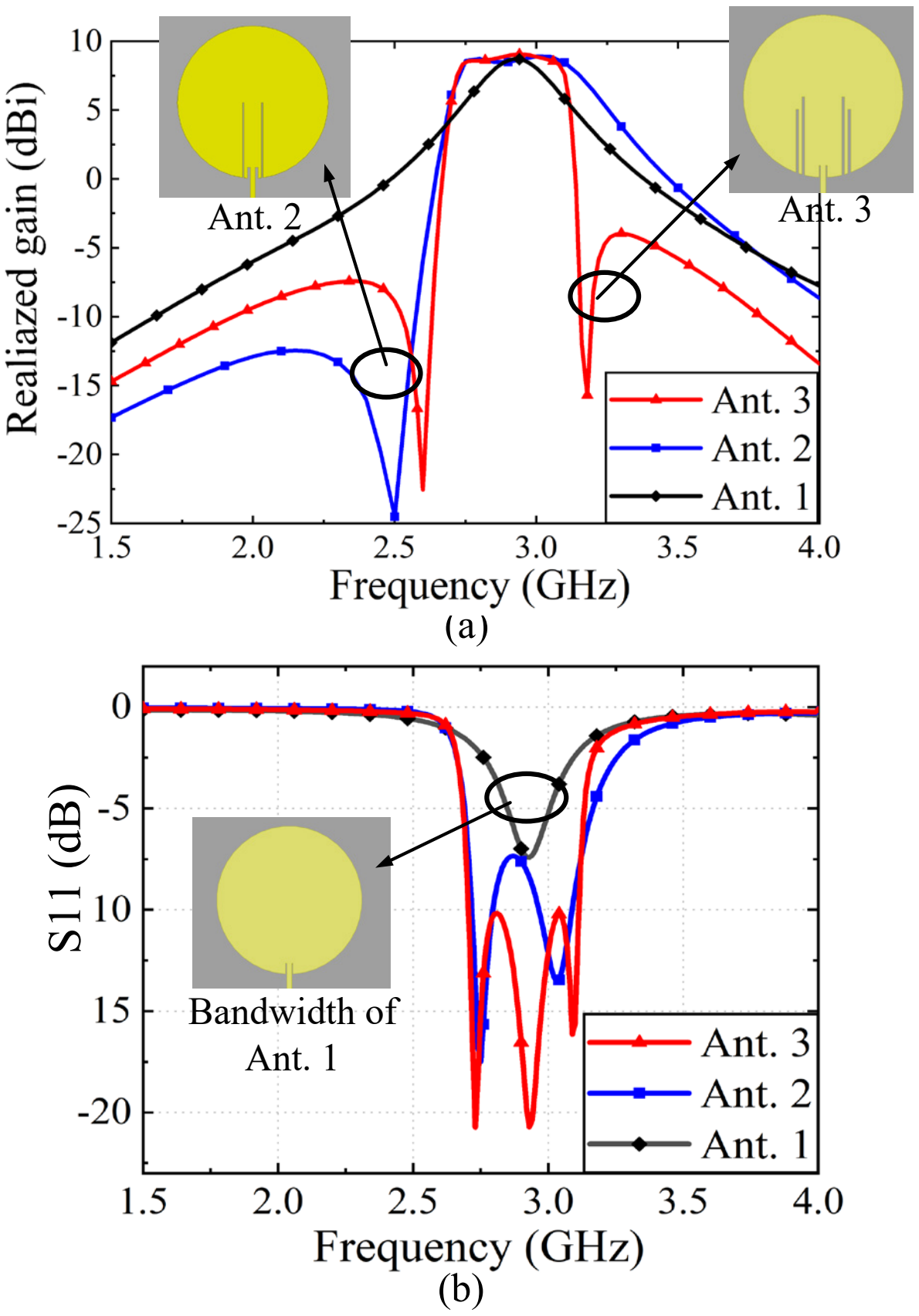

**Fig. 3.** Realized gain and S11 of the corresponding antenna structures with different pairs of slots.

*B. Design of Filtering Structure*

Based on the evolution of antenna structures, the realized gain and S11 results for Ant. 1 to Ant. 3 with different pairs of vertical slots are shown in Fig. 3. The realized gain results of different antenna structures are shown in Fig. 3(a). By attaching a pair of longer slots from Ant. 1 to Ant. 2, a radiation null is achieved at the lower band edge. Moreover, with adding a pair of shorter slots on the outside of longer slots, the other radiation null is produced at higher band edge. With two pairs of slots, Ant. 3 achieves a stable in-band gain of 8 dBi and an out-of-band radiation suppression more than 12 dB, demonstrating good filtering capability. Besides, the S11 of different antenna structures is shown in Fig. 3(b). It can be observed that after adding slots to the Ant. 1, the bandwidth is correspondingly broadened. Furthermore, the S11 in both band edges of Ant. 3 varies more rapidly compared with Ant. 1, which is caused by adding radiation nulls for filtering capability. Compared to the conventional antennas, the out-of-band gain of filtering antenna exhibits a more rapid and pronounced decline for effective out-of-band signal suppression. Besides, the operating frequency of the antenna can be scaled for different applications. Firstly, the operating band of the baseline structure (Ant. 1) can be effectively scaled by adjusting the radius of circular radiation patch. Then, by adjusting the frequencies of radiation nulls to align with the new operating band, two radiation nulls will locate at the lower and higher band edges of the new operating band, achieving effective out-of-band radiation suppression. The integrated filtering design achieves both broader bandwidth and effective out-of-band radiation suppression.

To further elucidate the radiation of Ant. 3, the E-field distributions and radiation patterns are shown in Fig. 4. Firstly, as illustrated in Fig. 4(a), the E-field distribution of Ant. 3 in the operating band is given. The upper and lower edges of the rectangular patch have the E-field distribution with the same phase, which makes the patch radiate as a conventional patch antenna. Therefore, Ant. 3 demonstrates a good directional radiation pattern in the operating band. Fig. 4(b) shows the E-field distribution and radiation pattern at the lower band radiation null. The main E-field distribution with the reverse phase is perpendicular to the longer slots. Consequently, the in-put energy fails to radiate effectively, leading to low radiation efficiency. Meanwhile, the radiation pattern of Ant. 3 splits at the center, significantly reducing its energy in the main radiation direction. Therefore, with low radiation efficiency and different radiation pattern out of band, a deep null with low realized gain responses at 2.6 GHz. The radiation at the higher band radiation null is similar to the radiation null working at 2.6 GHz. As shown in the Fig. 4(c), by attaching a pair of shorter slots working at the higher band edge, low radiation efficiency and weak energy in the beam direction can be observed, achieving effectively out-of-band radiation suppression.

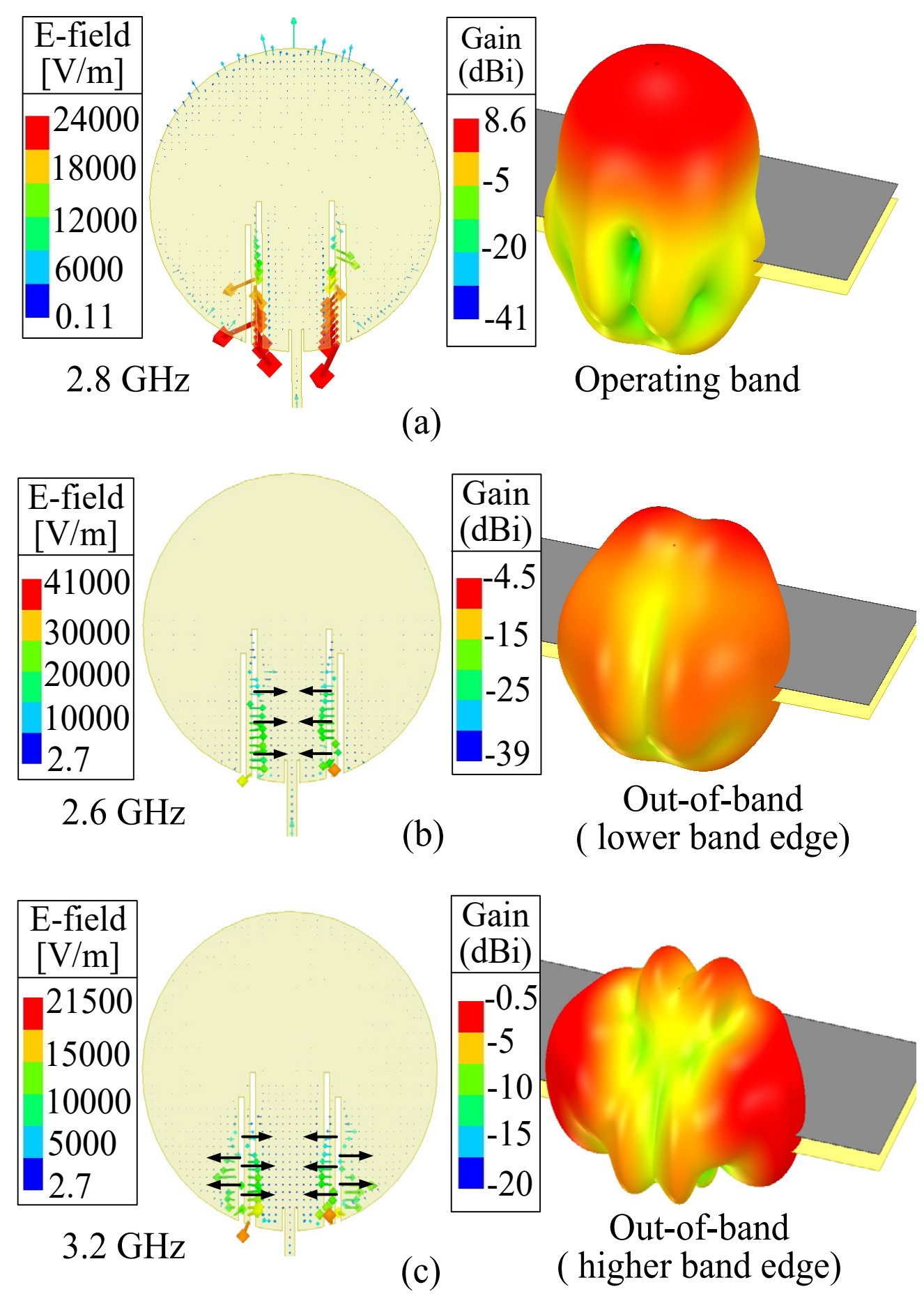

**Fig. 4.** E-field distributions and radiation patterns of Ant. 3. (a) operating band. (2.8 GHz) (b) radiation null at lower band edge. (2.6 GHz). (c) radiation null at higher band edge. (3.2 GHz).

To gain a comprehensive understanding of the antenna's filtering operation, we add a characteristic mode analysis (CMA) for Ant. 1-3, providing insights into their radiation and filtering responses from a modal perspective. The index of modal significance (MS) is used to help us analyze the main characteristic modes in 2 GHz - 3.5 GHz, and the index of modal weighting coefficient (MWC) is used to measure the contributions of different modes to the total radiation. A larger MWC indicates a greater impact on the total radiation, while a smaller MWC suggests a minimal excitation. The MS results for Ant. 1-3 is provided in Fig. 5. Besides, Fig. 6 illustrates the MWC values for each antenna, highlighting the main excited modes and their MWC nulls. As shown in the Fig. 5(a), there are two main modes for Ant. 1, which is named as Mode 1 and Mode 2. From the MWC results shown in the Fig. 6(a), the

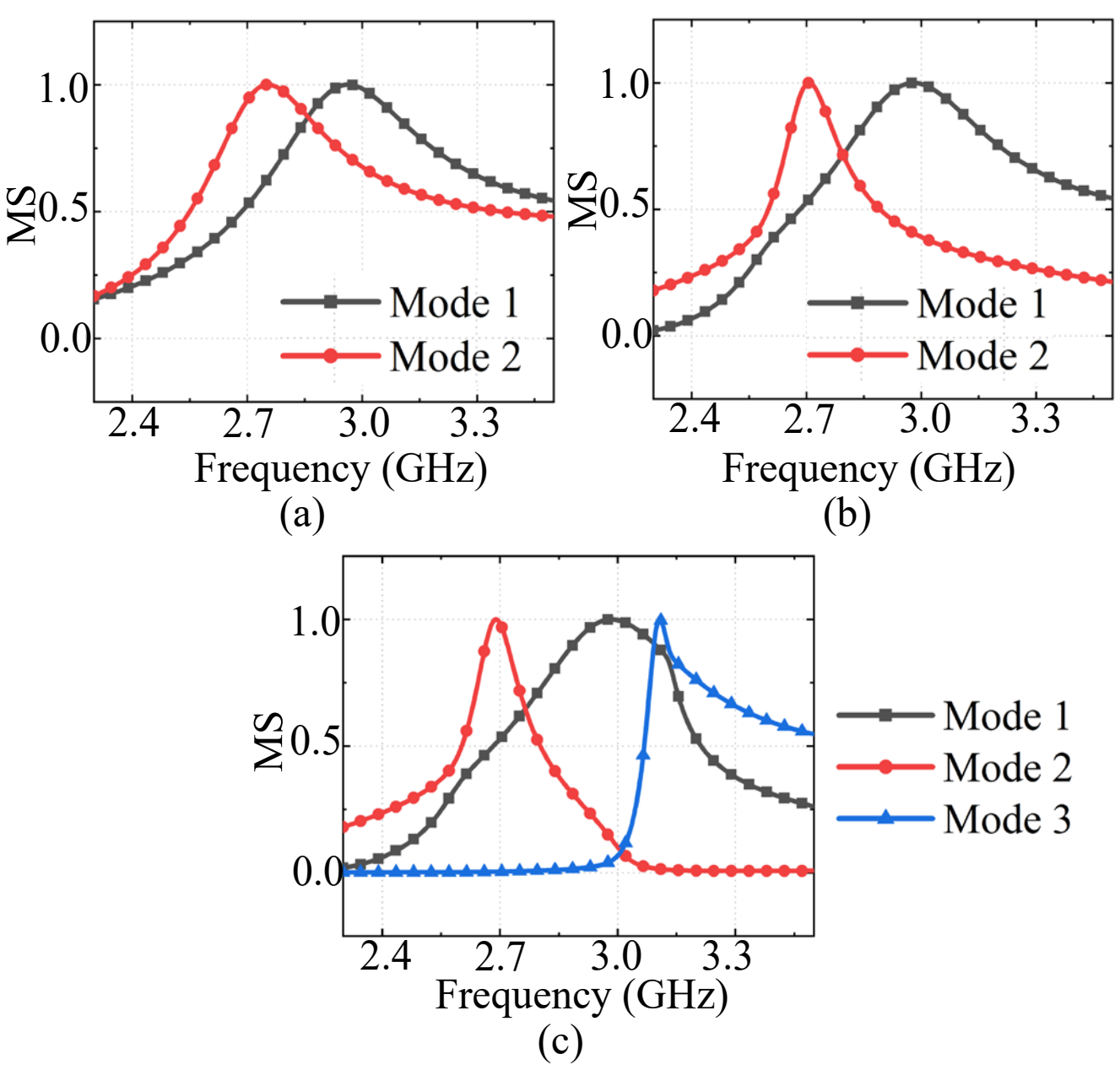

**Fig. 5.** Modal significance (MS) results for Ant. 1, Ant. 2, and Ant. 3. (a) Ant. 1. (b) Ant. 2. (c) Ant. 3.

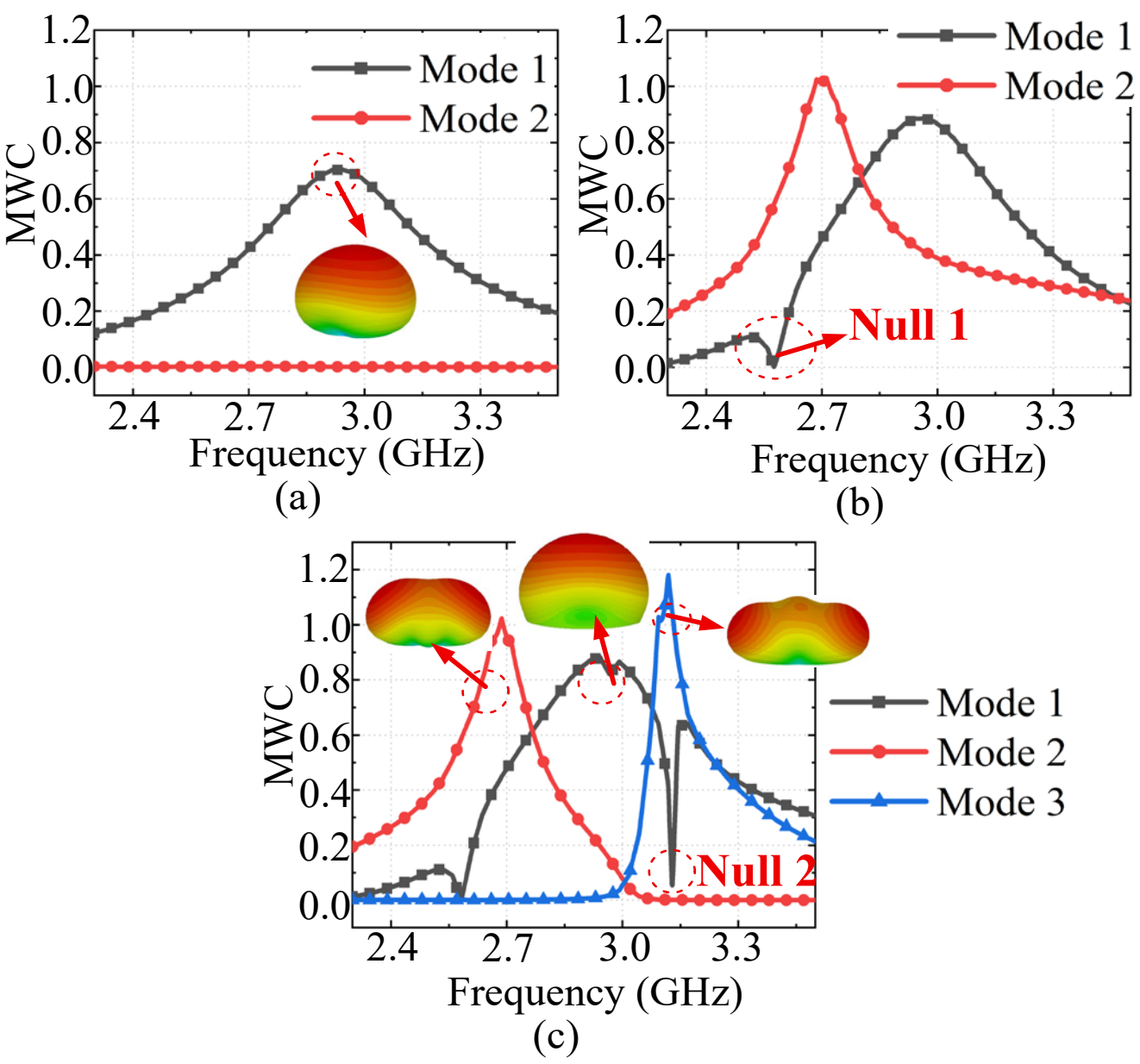

**Fig. 6.** MWC curves for excited modes of Ant. 1, Ant. 2, and Ant. 3, as well as radiation patterns at main frequencies. (a) Ant. 1. (b) Ant. 2. (c) Ant. 3.

MWC of Mode 2 is small, which mains the radiation of Ant. 1 is mainly excited by Mode 1. The Ant. 1 exhibits directional radiation in the operating band, which aligns with the radiation pattern of Mode 1. For Ant. 2, two modes are excited as shown in the Fig. 5(b). Moreover, as shown in the Fig. 6(b), the MWC of Mode 1 exhibits a null at the lower band, along with the Mode 2 excited at Null 1. As shown in the Fig. 6(c), the Ant, 3 mainly includes three modes. Besides, from MWC results, the Mode 1 is excited in the operating band. At the higher band, the MWC of Mode 1 exhibits the other null (Null 2). Furthermore, at the Null 1 and Null 2, the Mode 2 and Mode 3 are excited out of band, respectively. The out-of-band radiation patterns split at the center due to the unique field distribution of Mode 2 and Mode 3. This also agrees with the out-of-band radiation patterns shown in Fig. 4. With integrated filtering structure, Ant. 3 achieves filtering capability without any extra circuits in the planar state.

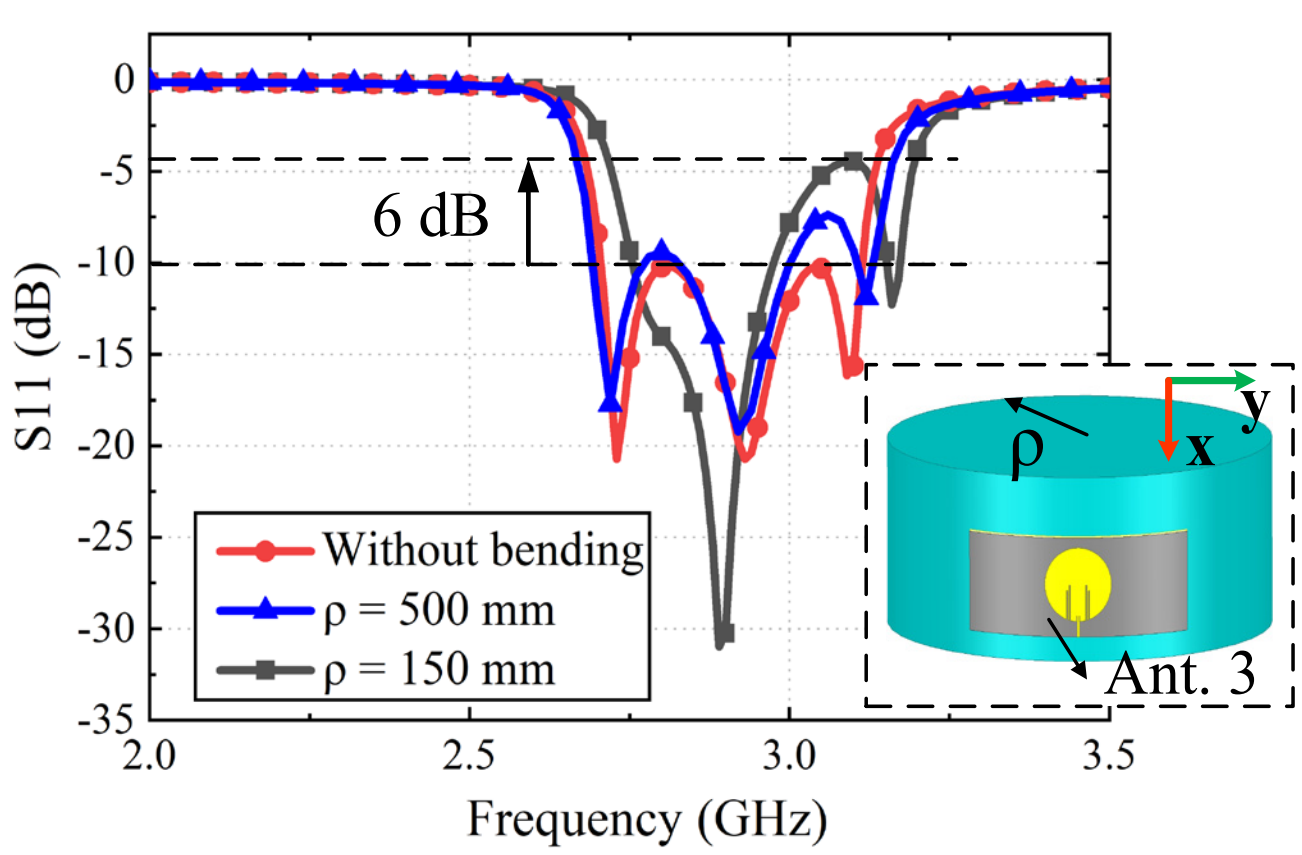

**Fig. 7.** Simulated S11 results of Ant. 3 before and after bending with different ρ.

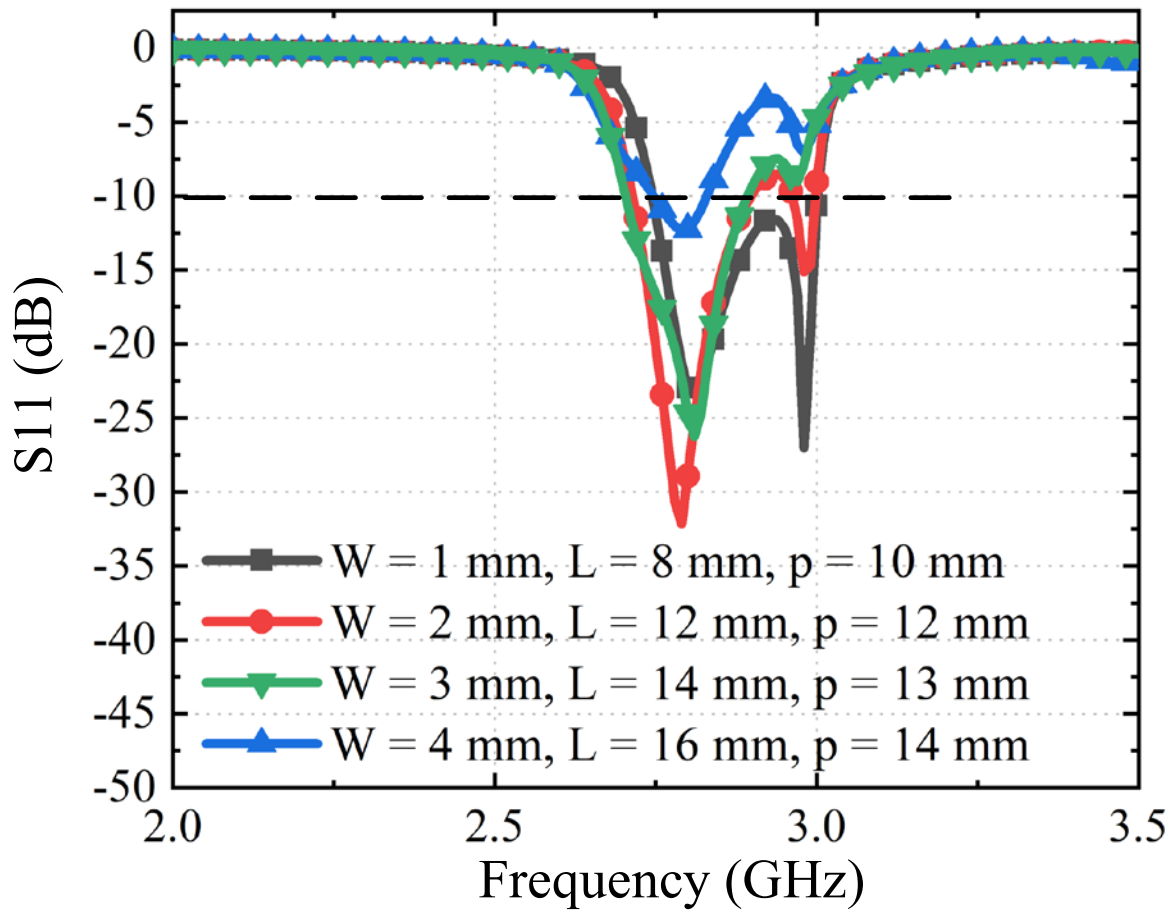

**Fig. 8**. Simulated S11 results of the antenna (ρ = 150 mm) with different width, length, and position of F-shaped slots.

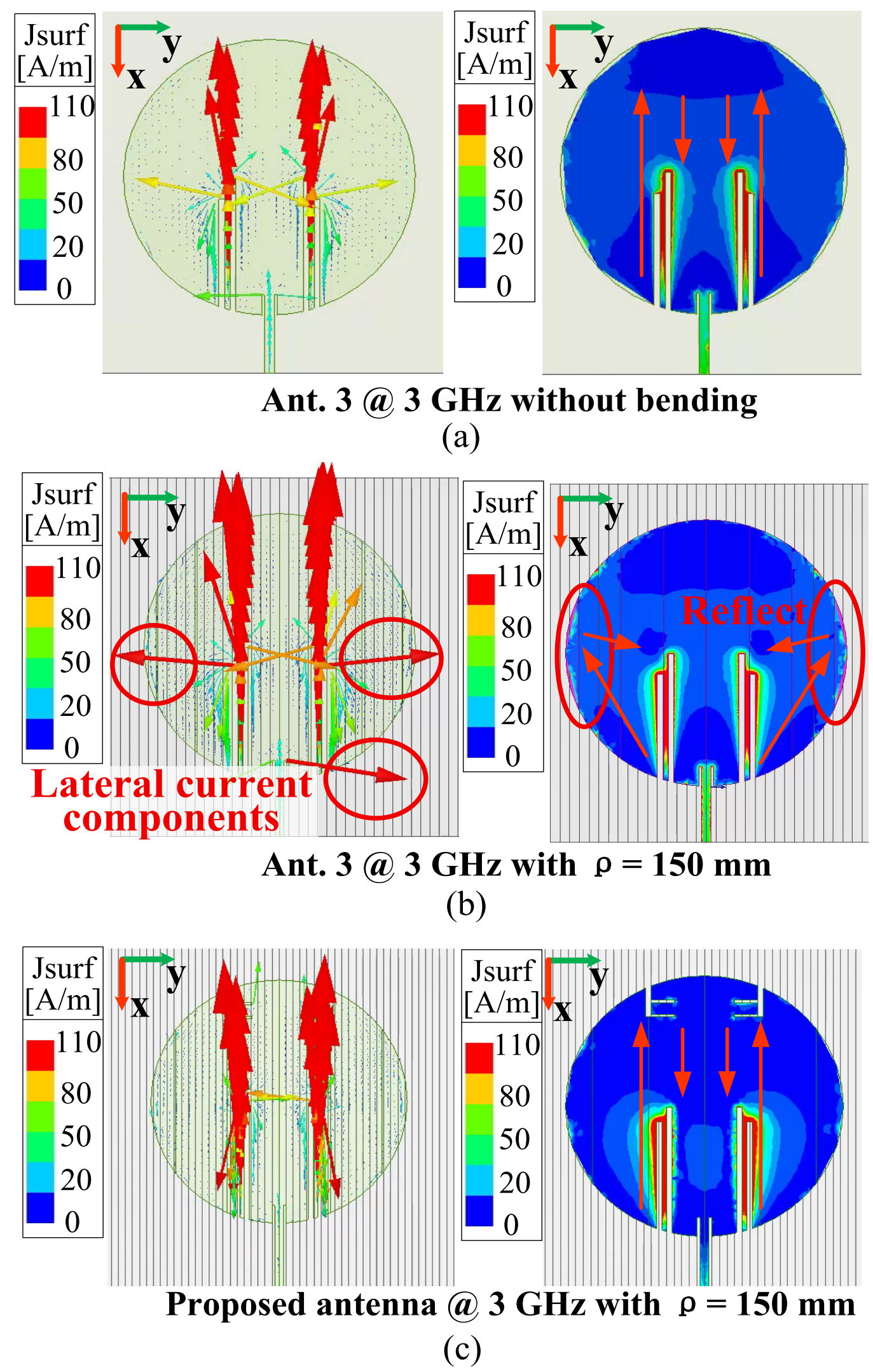

**Fig. 9.** Jsurf vector and magnitude distributions of flat Ant. 3, bent Ant. 3, and the proposed antenna after conformal with ρ = 150 mm at 3 GHz.

*C. Design for Stable Performance After Bending*

To validate the performance of Ant. 3 after bending, simulated S11 results with different bending radius are shown in Fig. 7. The planar Ant. 3 achieves great filtering capability and a −10 dB impedance bandwidth from 2.7 GHz to 3.2 GHz. However, after conforming the antenna to cylinder surface with ρ = 500 mm, the −10 dB impedance bandwidth of Ant. 3 narrows, which ranges from 2.7 GHz to 3 GHz. A degradation of S11 can be observed from 3 GHz to 3.2 GHz compared with unbending state. Besides, with increasing curvature, the S11 from 2.9 GHz to 3.2 GHz deteriorates from lower than -10 dB to about -4 dB when ρ = 150 mm. Correspondingly, the realized gain from 2.9 GHz to 3.2 GHz reduces due to the degraded reflection coefficient. The radiation of Ant. 3 deteriorates as bending curvature increases, which severely impacts its performance after flexible conformal.

To maintain stable performance after bending, a pair of inverted F-shaped slots are loaded as the structure evolving from Ant. 3 to the proposed antenna. Fig. 8 shows the simulated S11 results of the bent antenna (ρ = 150 mm) with different width, length, and position of F-shaped slots. It can be observed that with W = 1 mm, L = 8 mm, and p = 10 mm, the proposed antenna achieves stable −10 dB impedance bandwidth after bending. When the size of F-shaped slot is further enlarged, S11 results of the antenna deteriorates after bending. Therefore, the parameters of F-shaped slots are chosen as W = 1 mm, L = 8 mm, and p = 10 mm. To explore the principle behind how the F-shaped slot suppresses S11 deterioration, the Jsurf vector and magnitude distributions of the flat Ant. 3, bent Ant. 3, and the proposed antenna after conformal with ρ = 150 mm at 3 GHz are shown in Fig. 9. From Fig. 9(a), in the operating band, the main in-band current vector of flat Ant. 3 flows along X-axis with Mode 1 effectively excited in this state. As shown in Fig. 9(b), due to the bending on the Y-axis, a part of current flows laterally along the curvature direction, which weakens the Mode 1 resonance. These lateral current components are unable to radiate effectively. Therefore, the non-radiated lateral energy is reflected at the left and right sides of the antenna, resulting in an obvious degradation of reflection coefficient. To suppress lateral current components, a pair of inverted F-shaped slots are attached on the radiation patch. By introducing the F-shaped slots, the coupling of F-shaped slots effectively guides the main current to flow along X-axis after bending, enhancing the strength of Mode 1. Therefore, the lateral components can be suppressed on the proposed antenna effectively as shown in the Fig. 9(c), achieving stable S11, radiation pattern, and in-band gain under different curvature.

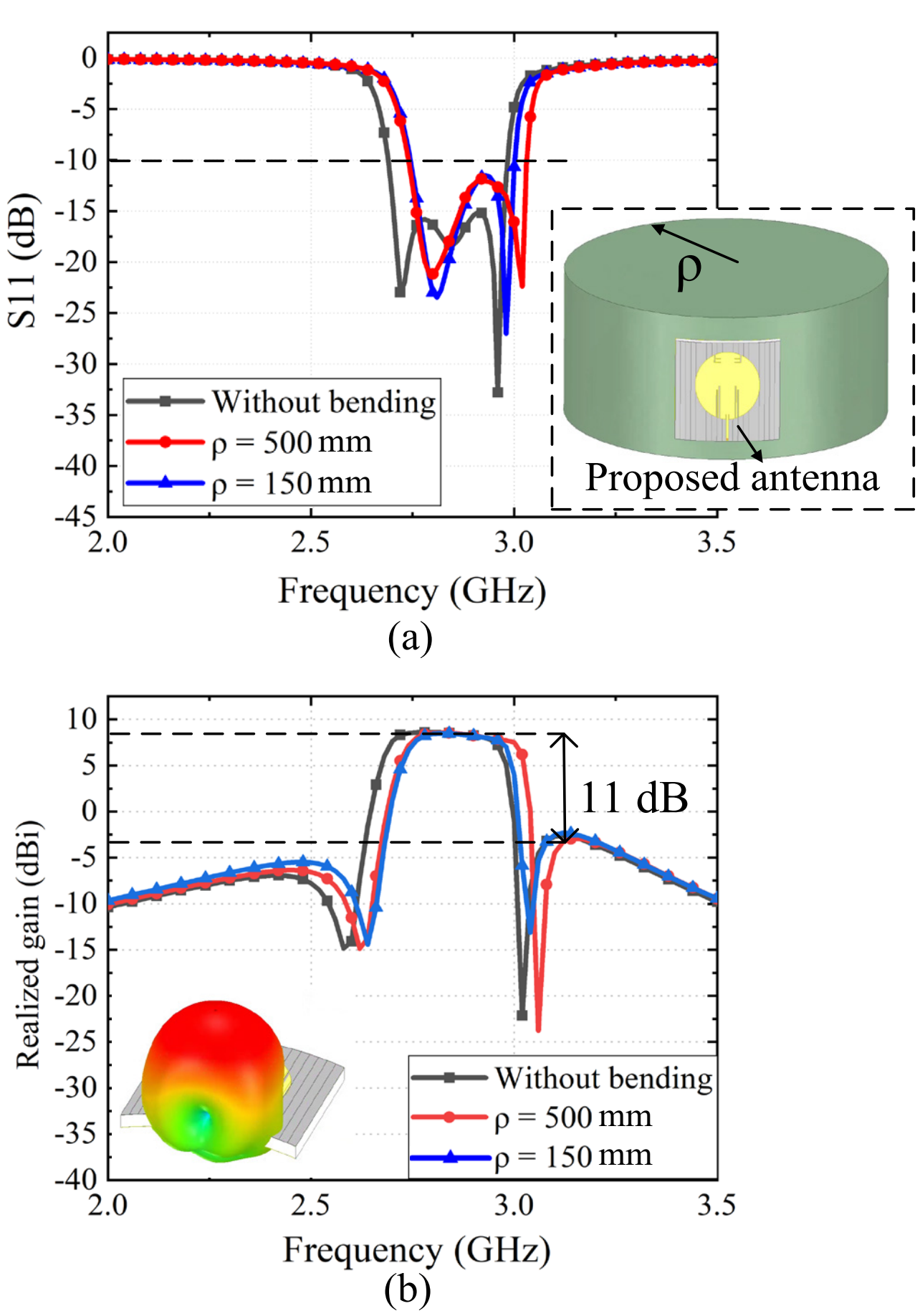

**Fig. 10.** Simulated S11 and realized gain results of the proposed antenna before and after bending. (a) S11 results. (b) realized gain results.

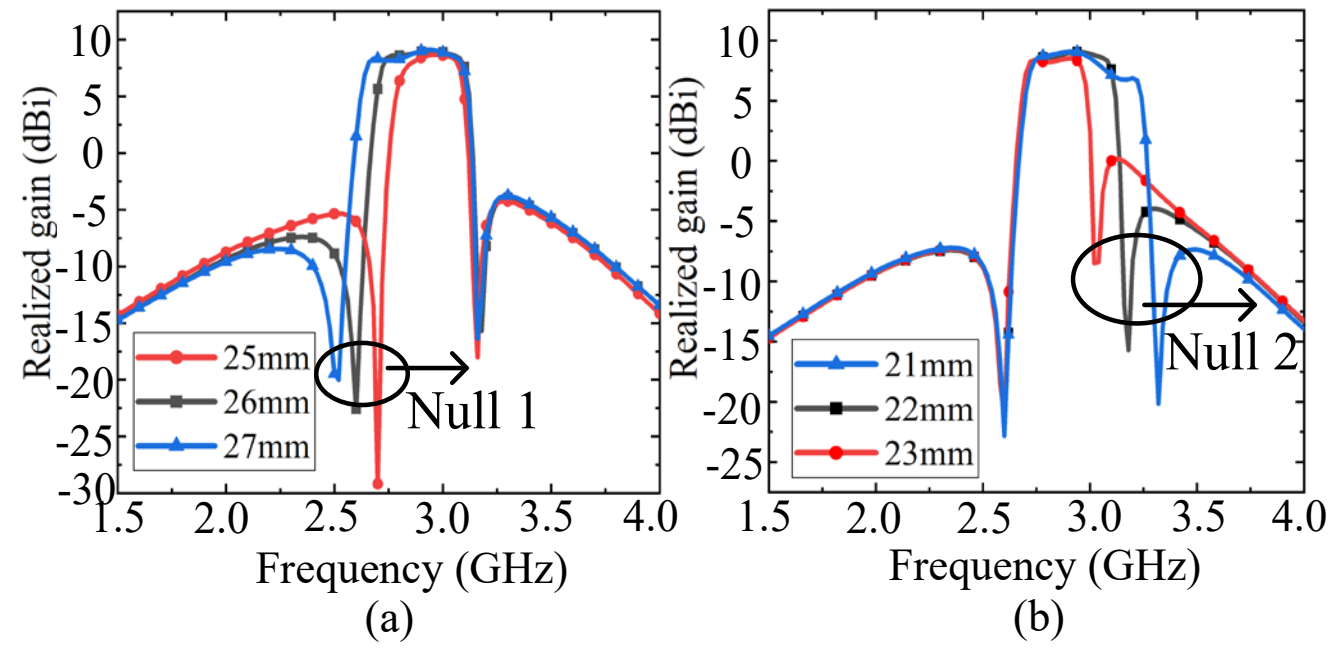

**Fig. 11.** Control of two radiation nulls (a) control of Null 1 with changing lengths of longer slots ($L_3$). (b) control of Null 2 with changing lengths of shorter slots ($L_4$).

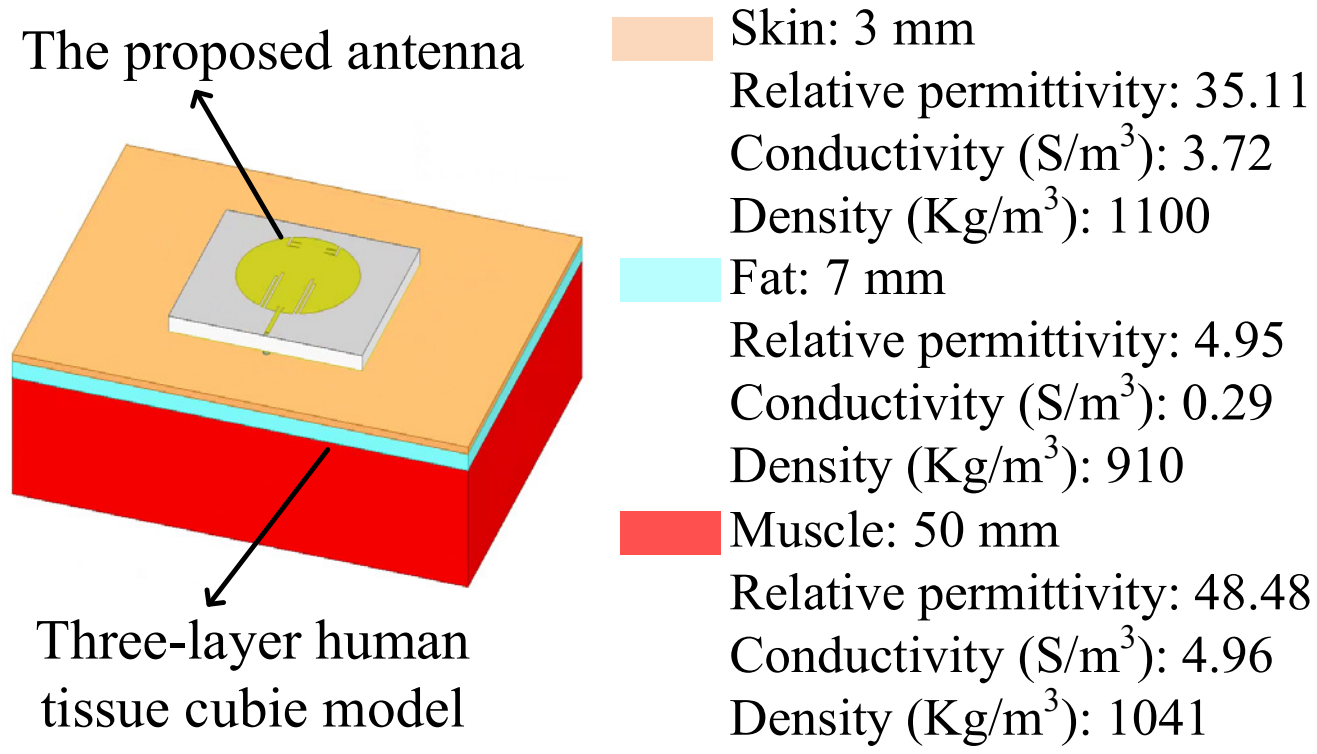

**Fig. 12.** Three-layer human tissue cubie model for wearable simulations.

The simulated realized gain and S11 results of the proposed antenna before and after bending are shown in Fig. 10. As shown in Fig. 10(a), after adding F-shaped slots, the bandwidth of the proposed antenna is 2.7 GHz to 3 GHz. As the curvature increases, the -10 dB impedance band of the proposed antenna remains stable. Moreover, as shown in Fig. 10(b), two radiation nulls can still position at both higher and lower band after bending, achieving stable out-of-band radiation suppression more than 11 dB under different curvature. Besides, in-band gain about 8.5 dBi with stable directional radiation pattern is realized. As discussed above, the proposed antenna achieves stable operating bandwidth, in-band gain and effective filtering capability after bending.

Fig. 11 shows the independent control of two radiation nulls with changing the lengths of slots. As shown in Fig. 11(a), by changing the length of the longer slots ($L_3$) from 27 mm to 25 mm, the frequency of Null 1 moves from 2.5 GHz to 2.7 GHz, while the other radiation null has almost no change. Besides, the control of radiation null at higher band (Null 2) is shown in Fig. 11(b). By adjusting the length of shorter slots ($L_4$) from 23 mm to 21 mm, the frequency of Null 2 is increased from 3 GHz to 3.2 GHz. The frequencies of two radiation nulls can be changed by adjusting the lengths of slots. Besides, the adjustment of one radiation null exhibits low influence on the frequency of the other radiation null, demonstrating excellent independent control capabilities.

### *D. Simulations of Antenna Performance After Wearing*

To further investigate the on-body radiation performance of the proposed antenna, joint simulations of the antenna and human tissue cubie model are carefully conducted. Fig. 12 shows the three-layer human tissue model for wearable simulations. The tissue model is consisted of the following layers in order, the skin tissue with a thickness of 3 mm, the fat tissue with a thickness of 7 mm, and the muscle tissue with a thickness of 50 mm. The thickness of each layer in the three-layer human body model is selected to represent typical torso values [33]. A 200 mm × 150 mm × 60 mm three-layered model is built in the HFSS simulator tool, where the antenna is separated from it by a gap distance of 10 mm. As shown in Fig. 13, the simulated realized gain and S11 results of the proposed antenna which is flexible conformal to the human tissue cubie model with different bending radius ρ are illustrated. As shown in Fig. 13(a), the proposed antenna has a −10 dB impedance bandwidth from 2.7 GHz to 3 GHz when attached to the human body model. Moreover, the operating band exhibits minimal variation with different bending radii. Furthermore, the realized gain results are shown in Fig. 13(b). It can be observed that the proposed antenna can achieve an out-of-band radiation suppression more than 15 dB in the lower band edge and exceeding 11 dB in the higher band edge after wearing. Besides, the proposed antenna maintains well directional radiation pattern and achieves an average in-band gain of 8.5 dBi. The simulated on-body performance of the proposed antenna keeps stable compared with results in the free space. This is primarily attributed to the antenna structure with full metal ground, which effectively mitigates the impact of human tissue on the performance of antenna. The antenna can maintain stable out-of-band radiation suppression, bandwidth, and gain when the bending radius exceeds 75 mm. In summary, the proposed antenna maintains consistent operating bandwidth and stable directional radiation after on-body bending. Furthermore, it demonstrates effective wearable filtering capabilities which can efficiently suppress out-of-band signal interference.

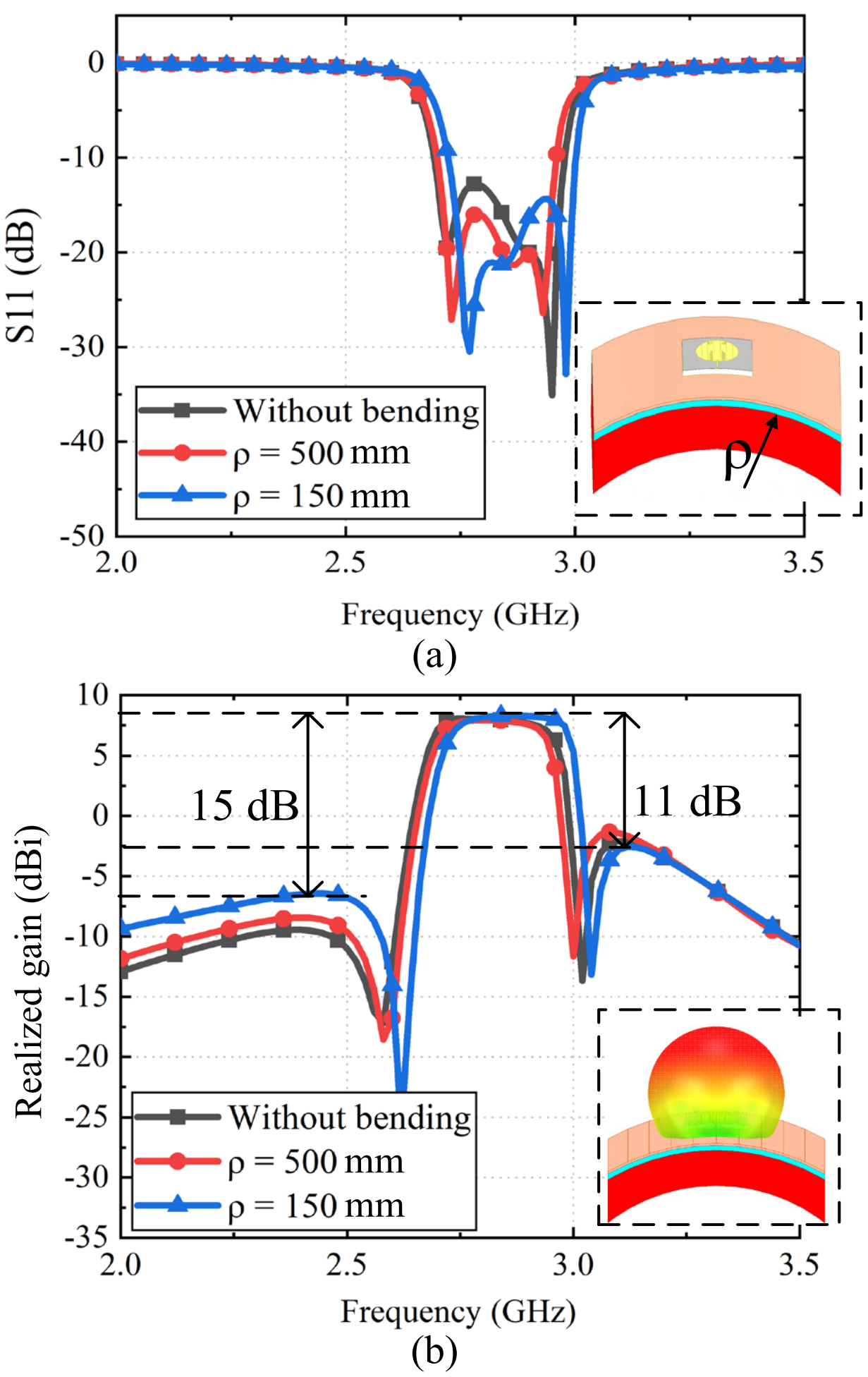

**Fig. 13.** Simulated results of the proposed antenna conformal to the human tissue cubie model with different bending radius. (a) S11 results. (b) realized gain and radiation pattern results.

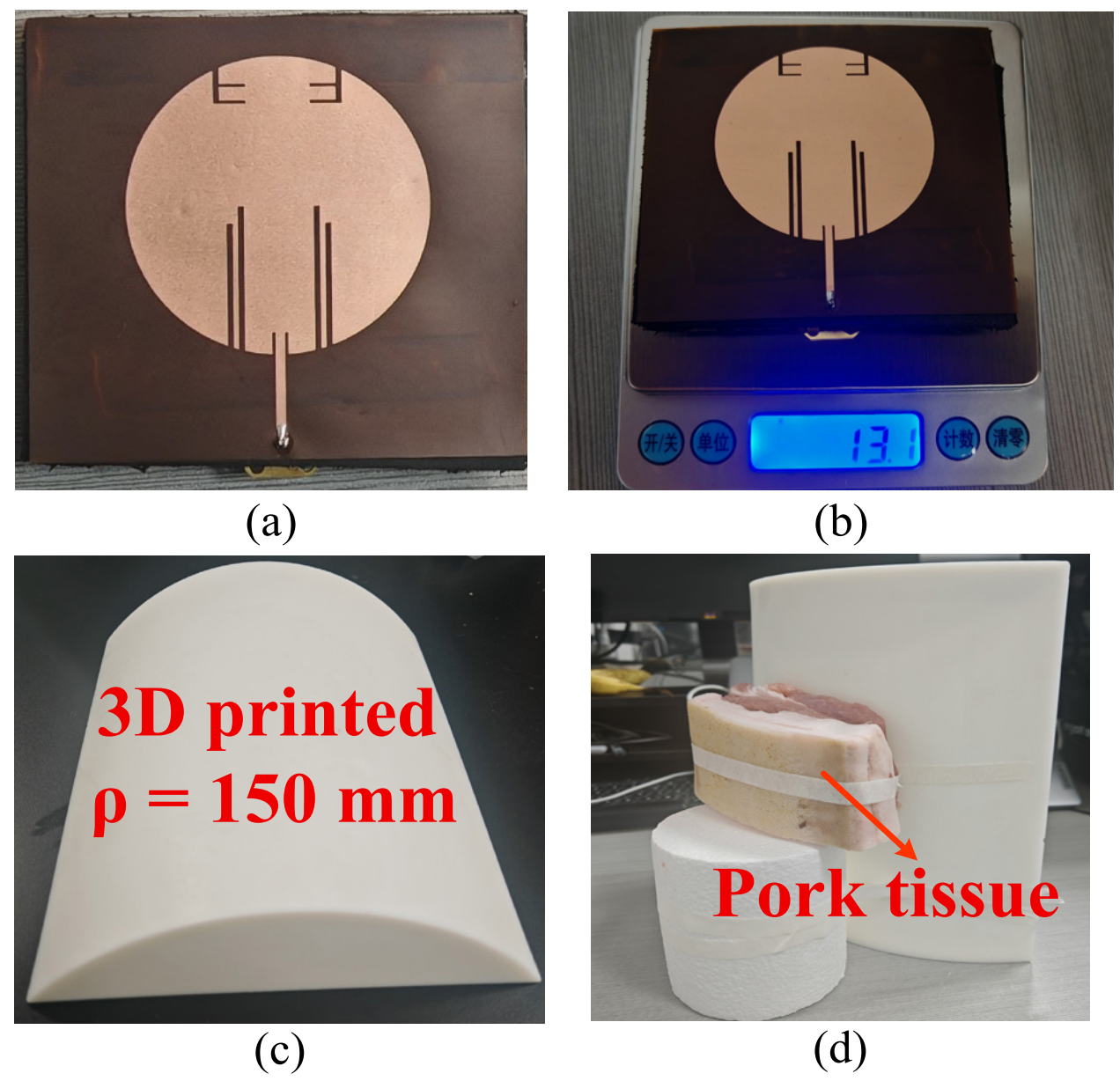

**Fig. 14.** Antenna prototype and simulations of curved skin model (a) antenna prototype. (b) weight measurement (SMA included). (c) 3D printed curved surface with ρ = 150 mm. (d) simulated curved human body model.

## III. Results And Discussion

### *A. Antenna Prototype and Structure of Curved Skin Model*

The proposed antenna is fabricated, and photograph of the prototype is shown in Fig. 14(a). As shown in Fig. 14(b), the weight of the fabricated antenna is precisely measured. The results reveal that the prototype is only 13.1 g (SMA included), which is a significant low value for weight-sensitive wearable applications. The SMA connector is chosen as SMA-KDF18, whose working frequency range covers DC - 18 GHz. The copper is printed on the PI substrate with a thickness of 0.5 OZ. The density of the foam has a certain impact on antenna performance. Therefore, foam with a relatively low density is selected during the fabrication process to ensure stable antenna performance. The curved model with a bending radius of 150 mm is depicted in Fig. 14(c), which is 3D printed with curable resin. To test the on-body performance, a pork trunk of approximately 180 mm × 100 mm × 21 mm is used. The pork trunk includes three layers: the skin, a fat layer and underlying muscle tissue. During on-body testing, the pork tissue is conformally attached to the 3D printed model to simulate curved human surface as shown in Fig. 14(d).

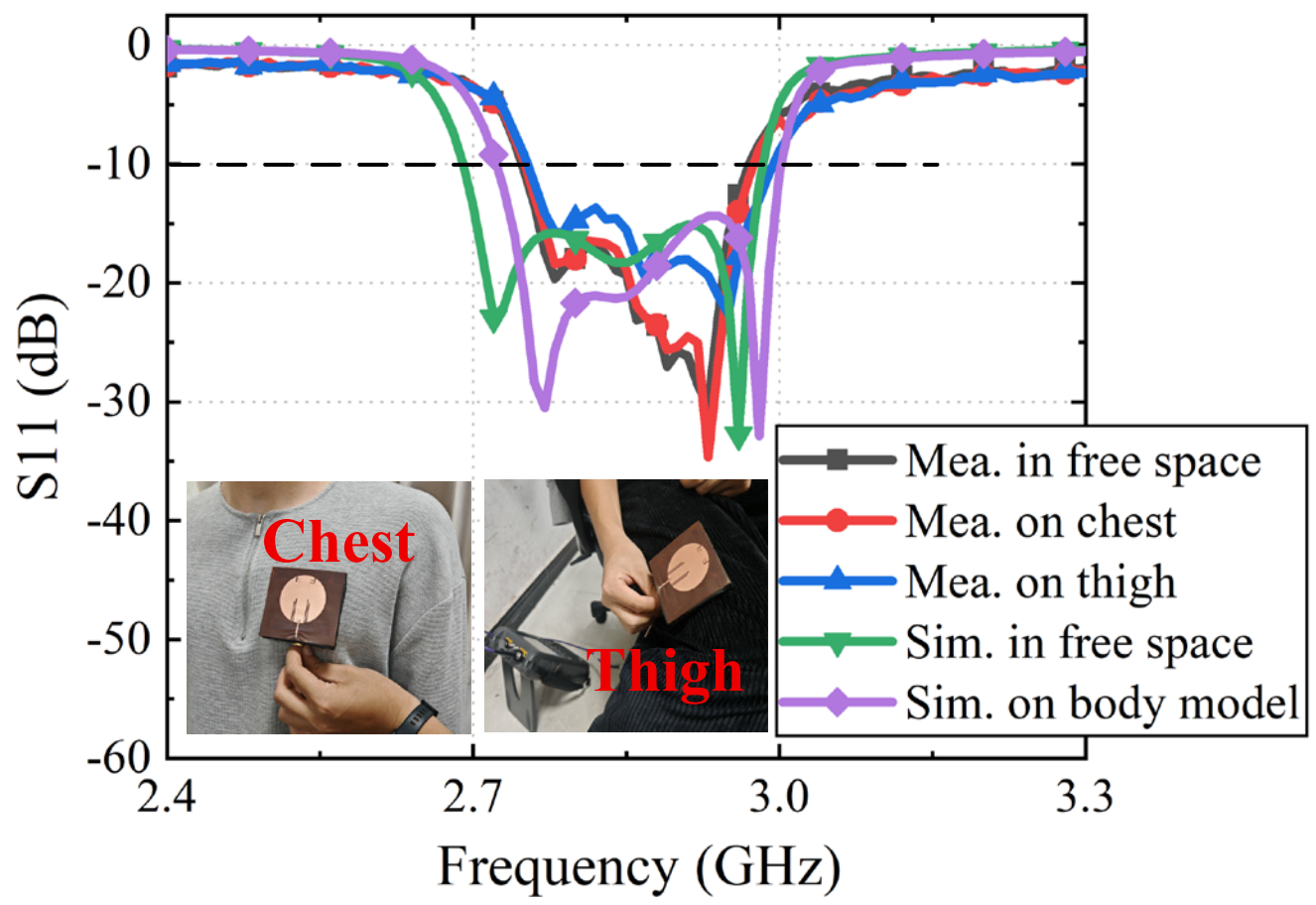

**Fig. 15.** Measured and simulated reflection coefficients in the free space and measured results on the chest and thigh.

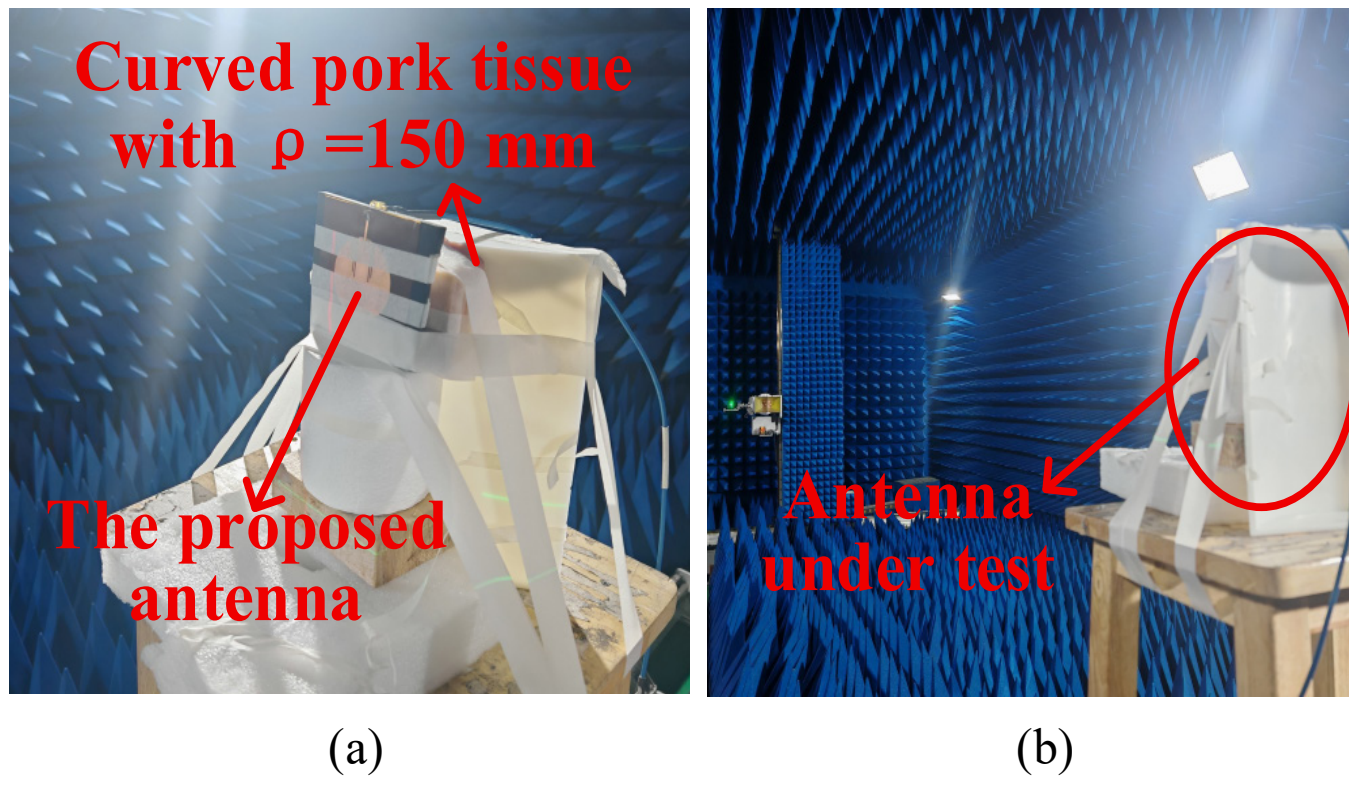

**Fig. 16.** Measurement set up and fabricated platform with antenna prototype. (a) antenna is attached to pork surface with cylindrical bending radius of 150 mm. (b) far-field testing in an anechoic chamber.

*B. Reflection Coefficients Measurement*

The experimental setup and measured S11 results are shown in the Fig. 15. To further validate the on-body performance, the reflection coefficients are measured by placing the proposed antenna in the free space and clinging to real human, including chest, and thigh. The measured −10 dB impedance bandwidth of the proposed antenna placed in free space is 300 MHz (2.7 – 3 GHz) with a fractional bandwidth of 10.5%, which is in good agreement with the simulated results. The measured results on the chest and thigh are presented as well. The antenna is conformally placed on real human tissue, with a layer of clothing in between. Measurements were repeated during on-body test. To reduce error, the average value is calculated from repeated measurements. The measured -10 dB impedance results show minimal error across repeated tests in the same scenario. This is primarily attributed to the antenna's whole ground structure and its design for stable performance under different bending conditions. Even with slight variations in wearing curvature and contact body area during repeated measurements, the measure results remain stable. Besides, the reflection coefficient is not significantly affected when the antenna is loaded on different body tissue as shown in the Fig. 15. Moreover, the bandwidth variations between free space results and on-body conditions are minimal, maintaining the stable -10 dB impedance bandwidth from 2.7 GHz to 3 GHz. The S11 results differ slightly when the antenna is conformally placed under different body curves. In addition, the simulated S11 on the human model is in a good agreement with the measured S11 on real human tissue, which verifies the reliability of the human model. In summary, the antenna preserves stable operating bandwidth both attached to the curved human body and in the free space. Moreover, if the antenna is bent in multiple directions, its filtering capability is expected to remain stable. Its in-band gain and S11 may experience a gradual, slight degradation with increased bending curvature.

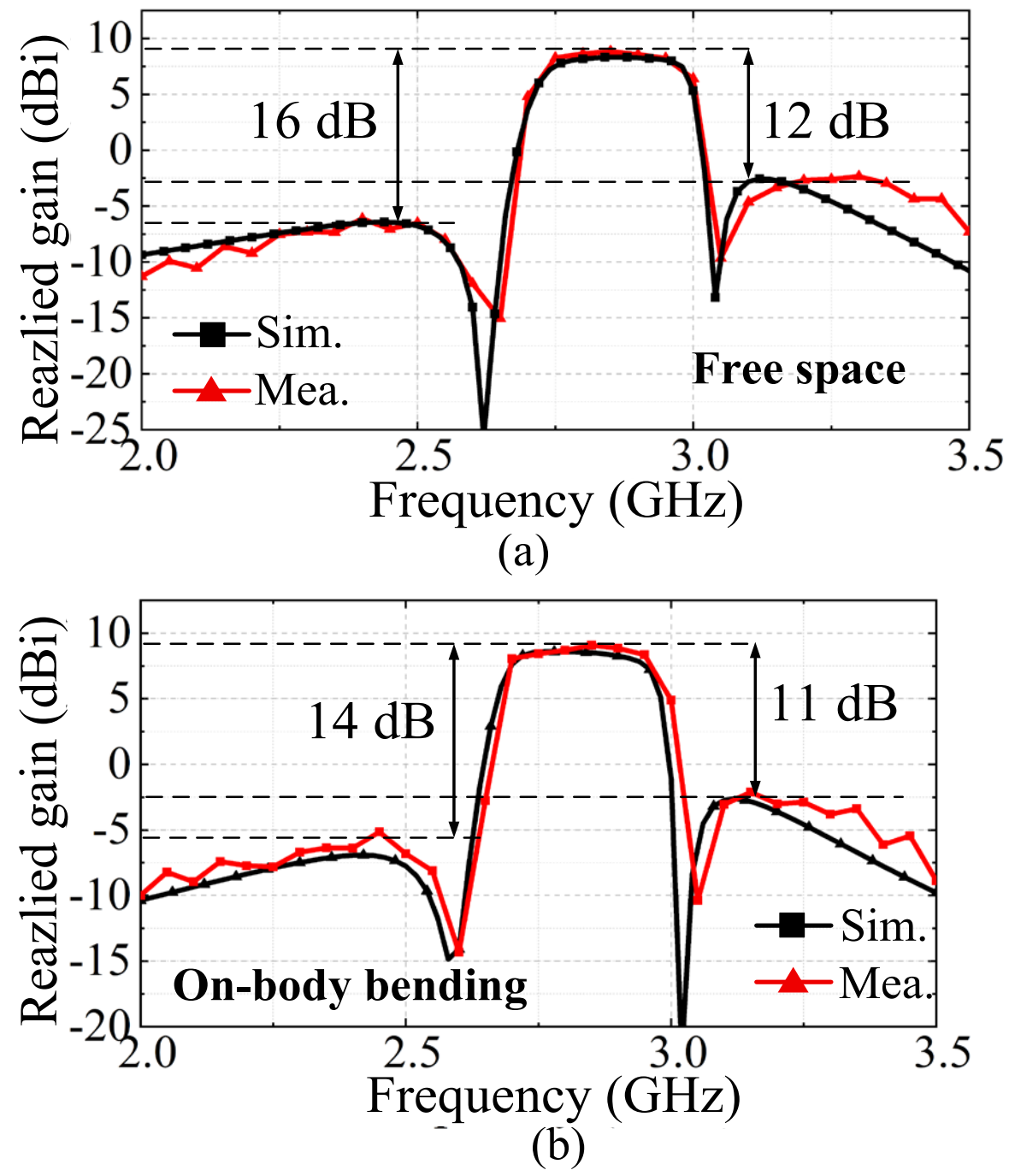

**Fig. 17.** Simulated and the measured realized gains for free space and on-body bending operation modes.

*C. Peak Gain and Radiation Patterns Measurement*

The proposed antenna is measured in two states. Fig. 16 presents the measurement set up and fabricated platform with antenna prototype. Firstly, the antenna is measured in the free space without bending. Besides, as shown in Fig. 16(a), a pork tissue is conformally attached to the 3D printed model with bending radius of 150 mm to simulate curved human body. Then, the proposed antenna is attached to the bent pork tissue, which can effectively simulate its working state in the wearable operation with combination effects of biological tissue and bending. Moreover, the far-field characteristics of two antenna states are measured in an anechoic chamber, as exhibited in Fig. 16(b).

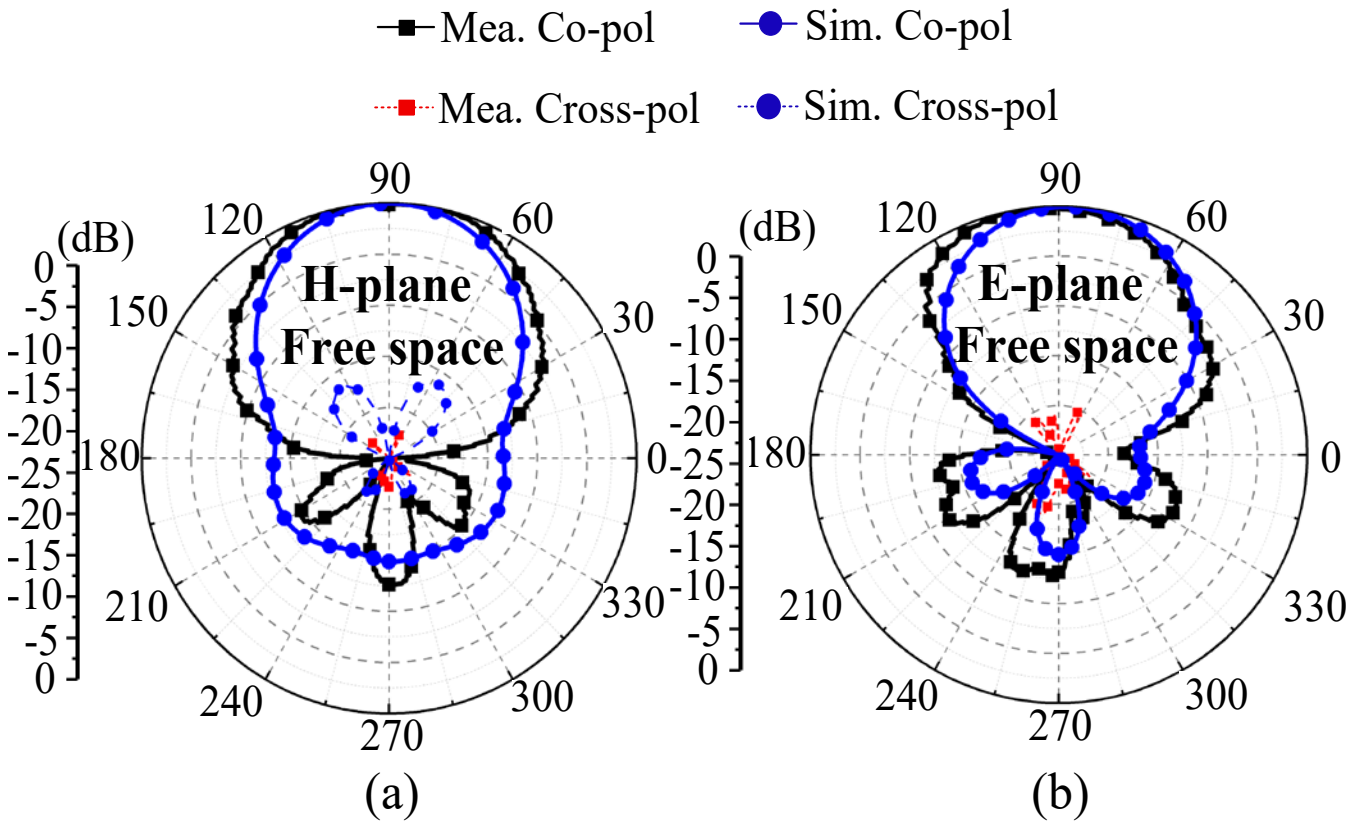

**Fig. 18.** Measured and simulated radiation patterns of the proposed antenna in the free space at 2.8 GHz.

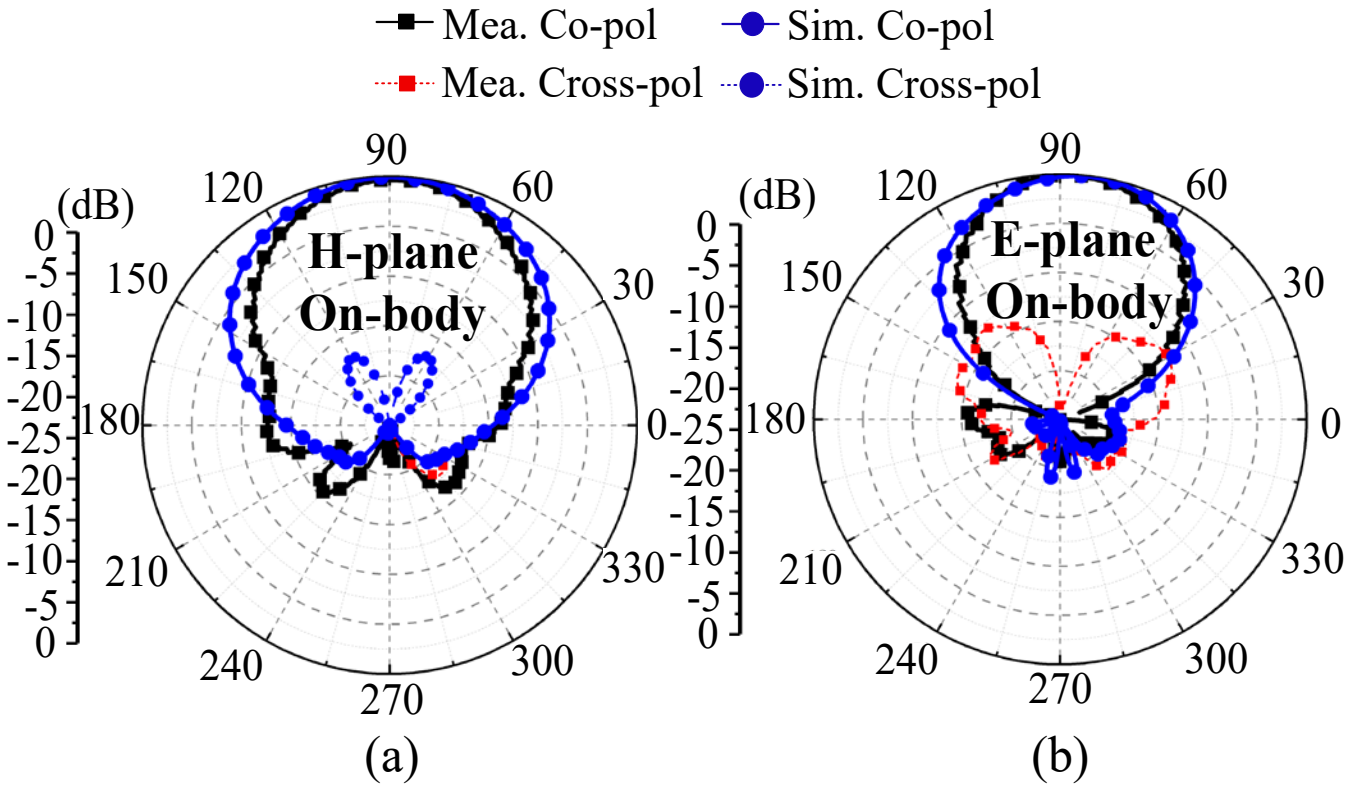

**Fig. 19.** Measured and simulated radiation patterns of the proposed antenna with on-body bending operation modes at 2.8 GHz.

The simulated and the measured realized gain results for free-space and on-body bending operation modes are shown in Fig. 17. The proposed antenna offers a peak gain of 9 dBi in the operating band in the free space. Besides, the antenna can achieve an out-of-band radiation suppression more than 12 dB in the higher band and 16 dB in the lower band, respectively. Moreover, under conformal with body tissue with ρ = 150 mm, the in-band gain of the antenna remains stable, which is about 8.5 dBi. Besides, the out-of-band radiation suppression is more than 11 dB in the higher band and 14 dB in the lower band, respectively. Some discrepancies are observed between measured and simulated results under the on-body operations. The differences can mainly attribute to the deviations in the thickness of biological tissue layers from the simulation model, along with assembly error during experimental model establishment. The measured and simulated results prove that the proposed antenna can achieve stable in-band gain and effective out-of-band radiation suppression under on-body bending operations and free space modes, which is well suitable for wearable IoT applications.

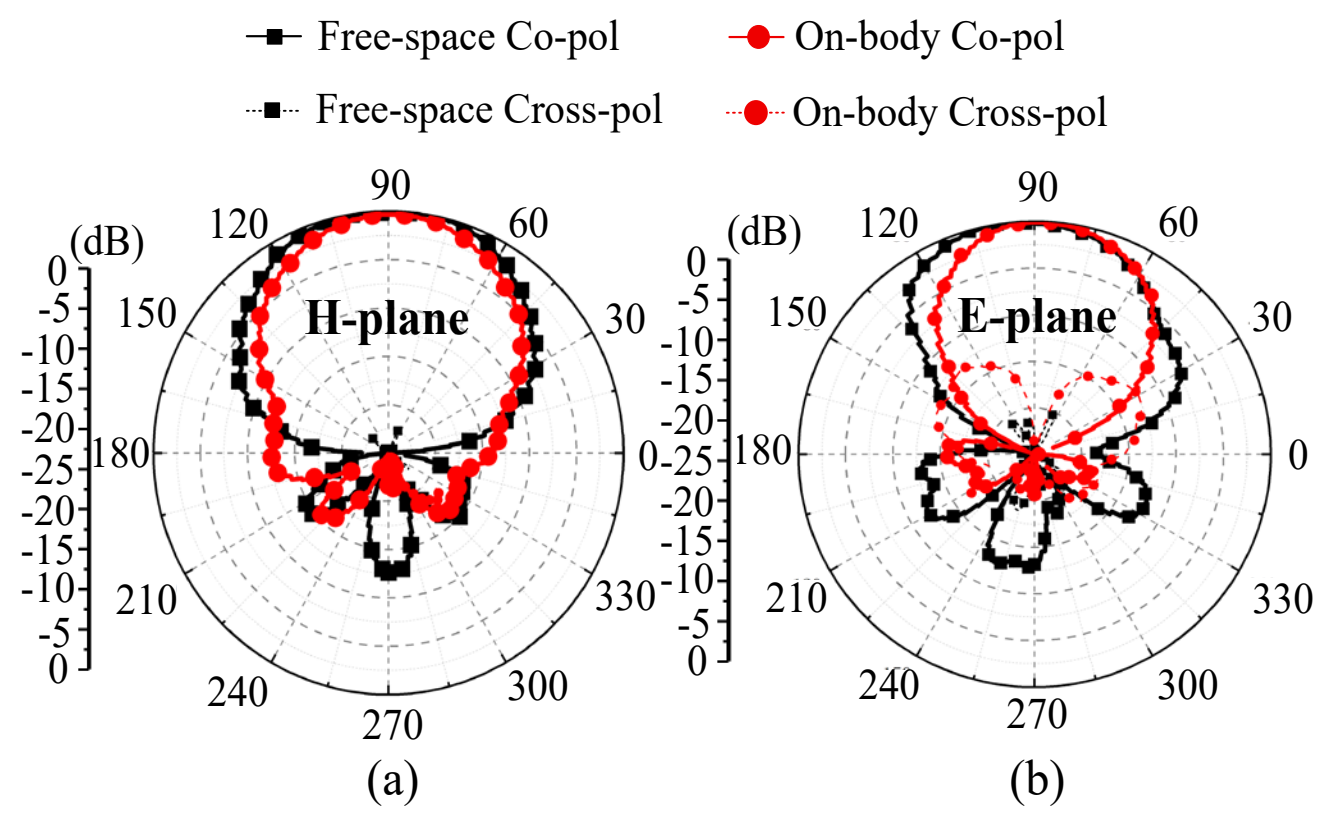

**Fig. 20.** Comparison of the measured on-body and free-space radiation patterns at 2.8 GHz.

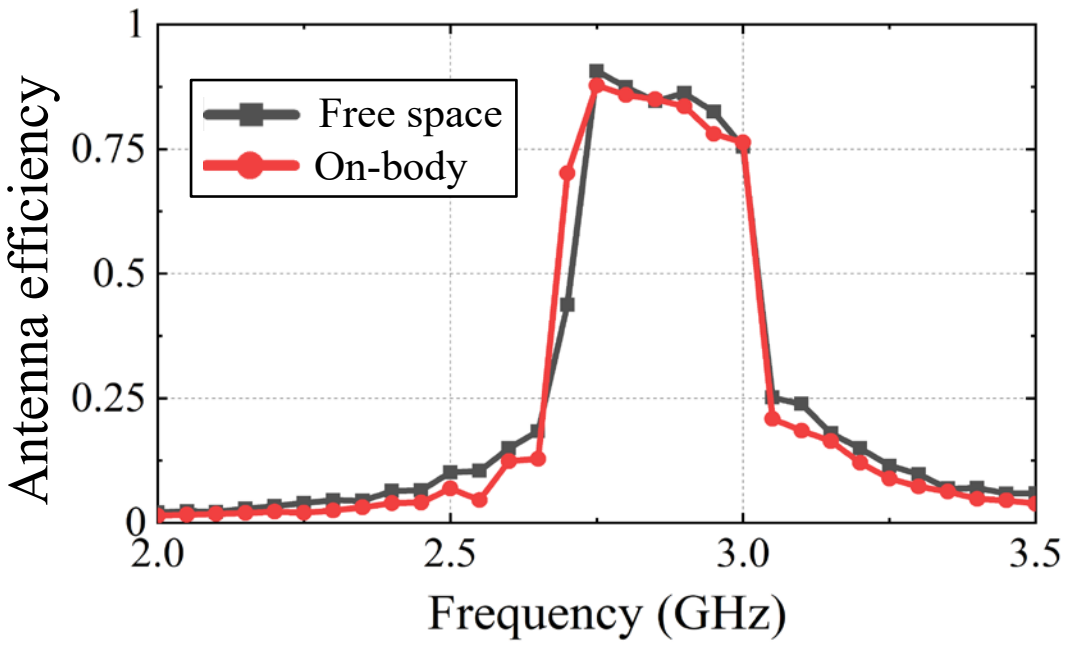

**Fig. 21.** Measured total antenna efficiency (mismatch included) of the proposed antenna for free space and on-body bending operation modes.

As shown in Fig. 18, the simulated and measured radiation patterns of the planar antenna in the free space at 2.8 GHz are depicted. It can be observed that the proposed antenna exhibits great directional radiation with front-to-back ratio (F/B) more than 15 dB. Moreover, the measured cross-polarization levels are all less than −20 dB. The simulated and measured radiation patterns of the proposed antenna under on-body bending operation modes at 2.8 GHz are given in Fig. 19. After flexible conforming, the measured F/B are all more than 20 dB, and measured cross-polarization levels are less than −20 dB. As shown in the Fig. 20, an improvement of F/B can be observed after being attached to the bent body tissue model. This development can mainly be attributed to the reflection from biological tissue and 3D printed model.

### *D. Antenna Efficiency and SAR Results*

Fig. 21 shows the total antenna efficiency that has been measured under free-space and on-body bending operation modes, including the impedance mismatch. With reference to the figure, the average measured total efficiency is higher than 75% in the passband. At both edges of the passband, the total efficiency drops rapidly, indicating high frequency selectivity with sharp roll-offs. In the lower and higher band edge, the peak efficiency is below 20%. The measured total antenna efficiency confirms that the radiation of the proposed antenna has been effectively suppressed in the stopband, as expected.

TABLE I
COMPARISON OF THE WORK WITH RELATED WORKS

| Ref. | Flexibility | Wearable Antenna | Filtering Ability | Filtering Circuit | Out-of-band suppression (dB) | Gain (dBi) | Weight | SAR (W/Kg) | Number of Nulls | Total Size ($\lambda_0 \times \lambda_0$) |
|---|---|---|---|---|---|---|---|---|---|---|
| [14] | No | Yes | No | None | None | 4.6 | None | 0.91 (10 g average 0.07 W input) | 0 | 1.4×0.88 (at 3.5 GHz) |
| [16] | Yes | Yes | No | None | None | 1.5 | None | 0.944 (1 g average 0.5 W input) | 0 | 0.385×0.37 (at 1.5 GHz) |
| [17] | Yes | Yes | No | None | None | 3.56 | None | 1.6 (10 g average None) | 0 | 0.44×0.44 (at 2.45 GHz) |
| [25] | No | No | Yes | Yes | >10.3 | 7.5 | None | None | 2 | 0.67×0.67 (at 3.5 GHz) |
| [26] | No | No | Yes | None | >10 | 2.4 | None | None | 2 | 0.4×0.006 (at 0.7 GHz) |
| [30] | Yes | No | Yes | None | >10 | 4.5 | None | None | 4 | 0.56×0.56 (at 1.4 GHz) |
| [31] | No | Yes | Yes | Yes | None | 5.2 | None | 1.6 (1 g average 0.67 W input) | 2 | 0.53 × 0.53 (at 4 GHz) |
| **This Work** | **Yes** | **Yes** | **Yes** | **None** | **>11** | **8.5** | **13.1 g** | **0.269 (10 g average 0.5 W input)** | **2** | **0.71×0.79 (at 3 GHz)** |

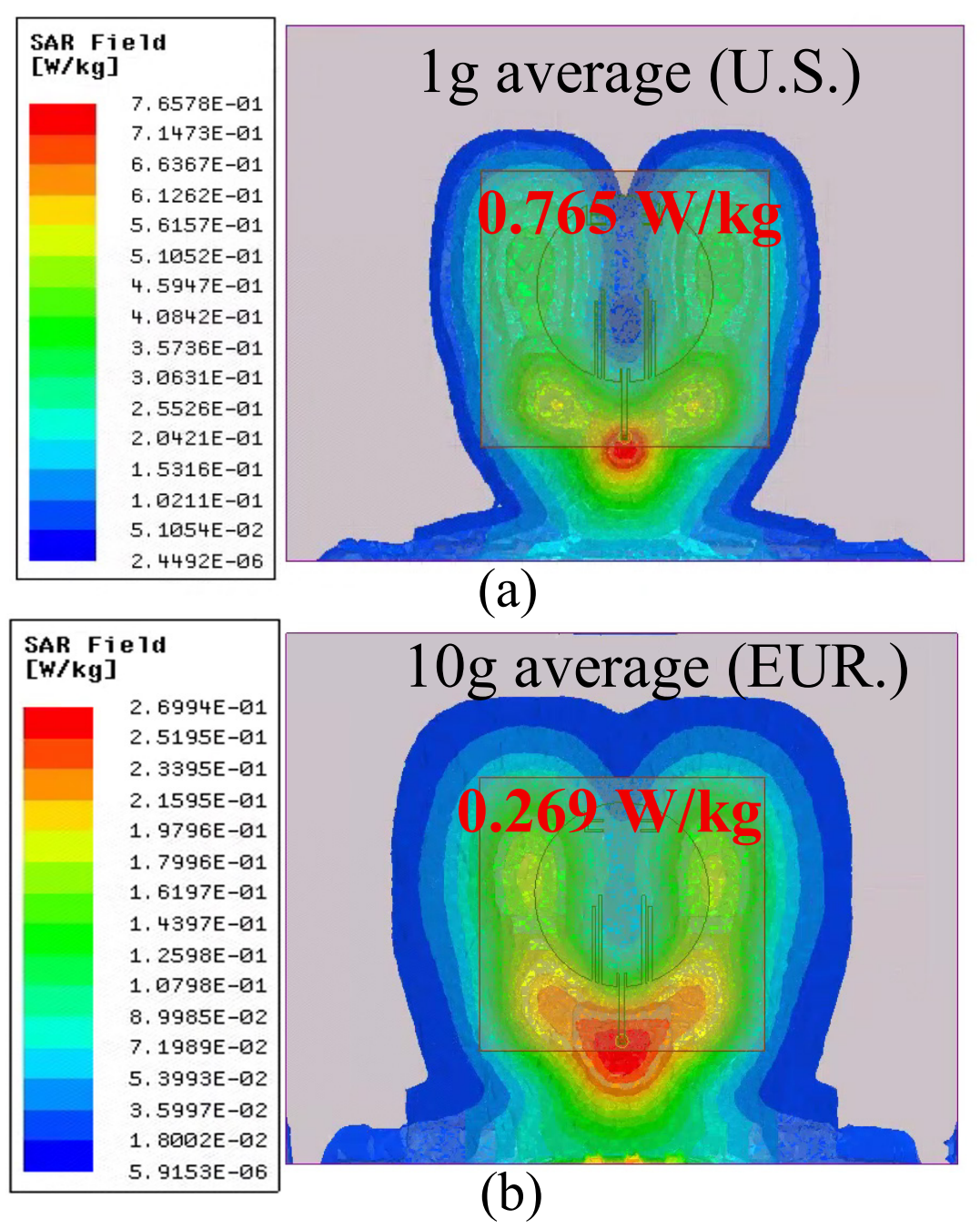

**Fig. 22.** SAR results of the proposed antenna at 2.8 GHz. (a) 1 g average. (b) 10 g average.

In addition to well radiation performance and out-of-band radiation suppression after wearing, SAR value is a crucial parameter to consider for the human well-being. The SAR value should not exceed predefined limits in order for the antenna to be approved as wearable. One of the two limits is the American standard, which is Federal Communications Commission (FCC) permitted, where the threshold is 1.6 W/kg averaged over 1 g of tissue. The second is the EU standard, International Electrotechnical Commission (IEC) permitted, where the threshold is 2 W/kg averaged over 10 g of tissue. The SAR distributions of the proposed antenna at 2.8 GHz under 1 g average and 10 g average are presented in Fig. 22. The initial mesh size in the HFSS is set to be less than 1/3 wavelength. The maximum refinement per pass is 30 %. The distance between the air box and the model is 30 mm, which is set in the range of λ/4 - λ/2 at the center frequency. Moreover, the SAR values were calculated based on the IEEE C95.1-2005 standard and the EN 50361-2001 standard with an input power of 0.5 W. The results show that the 1 g and 10 g averaged SARs at 2.8 GHz are less than 0.765 and 0.269 W/kg, which are far below the limits of 1.6 W/kg under the U.S. average and 2.0 W/kg under the EUR average, respectively. Moreover, the SAR value increases linearly with input power. When the input power increases to 1 W, the SAR results are still under relevant safety standards. Notably, the low SAR performance is attributed to its whole metal ground structure, which effectively reduces backward radiation toward human body during wearable operation.

### E. Comparison and Discussion

In Table I, a comprehensive comparison of the proposed antenna with related published works is summarized. From the table, it can be observed that none of the previously reported wearable filtering antenna has flexible functionality. Our flexible antenna maintains stable and effective filtering capability when attached to the curved body tissue, enabling seamless applications in wearable scenarios. Besides, filtering circuits are not applied in the proposed antenna design for filtering capability, which significantly mitigates insertion loss and develop in-band gain. Moreover, the existing flexible filtering antennas are not suitable for wearable applications because they cannot achieve low SAR and stable on-body bending performance. The proposed antenna achieves low SAR and stable performance after wearing, which is well suitable for wearable applications. Furthermore, lightweight and flexible substrates are employed, achieving a measured weight of 13.1 g (SMA included). From the above comparison, the proposed wearable antenna design offers distinct advantages: flexible wearability, enabling reliable filtering performance on curved human body surfaces, along with lightweight characteristics.

## IV. Conclusion

In this paper, a flexible and lightweight filtering wearable antenna with stable on-body performance is presented. The filtering capability is realized by introducing two pairs of vertical slots to the radiation patch. The design technique of stable performance after bending is explained by loading a pair of F-shaped slots to suppress non-radiated lateral current components along the curvature direction. The use of flexible substrates not only minimizes the antenna's weight but also facilitates its applications on the curved human body surfaces. The proposed antenna underwent fabrication and experimental validation, exhibiting superior out-of-band signal suppression capability. Furthermore, its electromagnetic properties remain stable even after loaded on different body locations with various curvature. These features collectively make it a promising candidate for wearable IoT applications.

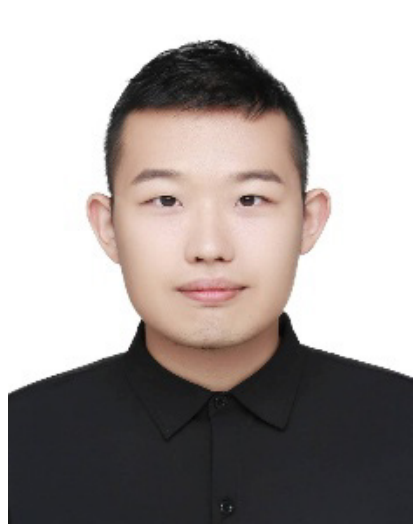

**Runkai Song** (Student Member, IEEE) was born in Qingdao, China. He received the B.S. degree in telecommunication engineering from Hainan University, Haikou, China, in 2023. He is currently pursuing the Ph.D. degree with Xidian University, Xi’an, China.

His research interests include filtering antennas, flexible antennas, and wearable antennas.

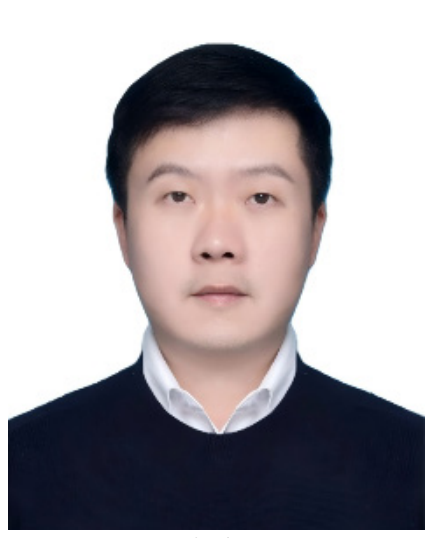

**Fan Qin** (Member, IEEE) received the B.S. degree in electronic information engineering and the Ph.D. degree in electromagnetic wave and microwave technology from Northwestern Polytechnical University, Xi’an, China, in 2010 and 2016, respectively. He was a visiting scholar with the University of Kent, Canterbury, UK, from 2013 to 2015. He joined the School of Telecommunications Engineering, Xidian University, Xi’an, China, in 2016. Currently, he is an Associate Professor with Xidian University. His research interests involve metamaterial antennas, millimeter wave antennas, vortex wave antennas, and metasurface-assisted wireless communication.

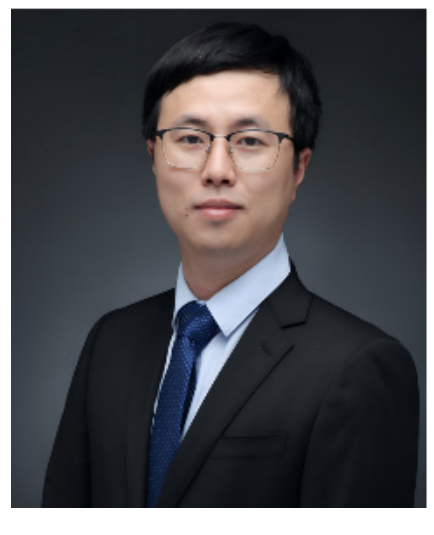

**Wenchi Cheng** (Senior Member, IEEE) received the B.S. and Ph.D. degrees in telecommunication engineering from Xidian University, Xian, China, in 2008 and 2013, respectively. He was a Visiting Scholar with the Department of Electrical and Computer Engineering, Texas A&M University, College Station, TX, USA, from 2010 to 2011. He is currently a Full Professor with Xidian University. He has published more than 200 international journal and conference papers in *IEEE Journal on Selected Areas in Communications*, *IEEE Magazines, IEEE TRANSACTIONS, IEEE INFOCOM, GLOBECOM,* and *ICC.* His current research interests include 6G wireless networks, electromagnetic-based wireless communications, and emergency wireless information. He received the IEEE ComSoc Asia–Pacific Outstanding Young Researcher Award in 2021, the URSI Young Scientist Award in 2019, the Young Elite Scientist Award of CAST, and four IEEE journal/conference best papers. He has served or serving as the Wireless Communications Symposium Co-Chair for IEEE ICC 2022 and IEEE GLOBECOM 2020, the Publicity Chair for IEEE ICC 2019, the Next Generation Networks Symposium Chair for IEEE ICCC 2019, and the Workshop Chair for IEEE ICC 2019/IEEE GLOBECOM 2019/INFOCOM 2020 Workshop on Intelligent Wireless Emergency Communications Networks. He has served or serving as the ComSoc Representative for *IEEE Public Safety Technology Initiative*, Editor for *IEEE Transactions on Wireless Communications, IEEE System Journal, IEEE Communications Letters, and IEEE Wireless Communications Letters.*

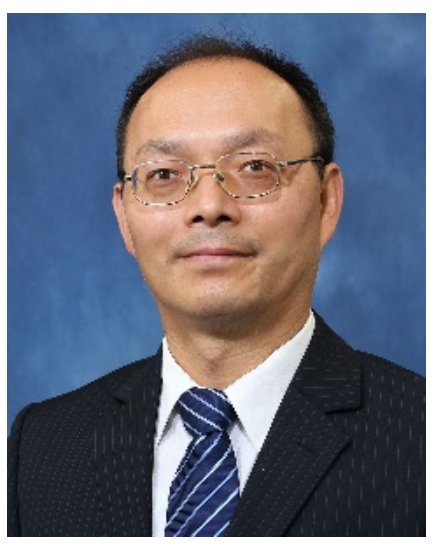

**Steven Gao** (M’01-SM’16-F’19) received the PhD from Shanghai University, China. He is a Professor at Department of Electronic Engineering, The Chinese University of Hong Kong, Hong Kong, where he is also the Director of Center for intelligent Electromagnetic Systems. Prior to joining CUHK, he was a Professor and Chair of RF/Microwave Engineering at the University of Kent (UKC), UK during 2013-2022, and became an Honorary Professor at UKC since 2022.

His research covers smart antennas, phased arrays, MIMO, reconfigurable antennas, broadband/multiband antennas, satellite antennas, and wireless systems (mobile and satellite communications, synthetic-aperture radars, IOT). He co-authored/co-edited 3 books (Space Antenna Handbook, Wiley, 2012; Circularly Polarized Antennas, IEEE & Wiley, 2014; Low-Cost Smart Antennas, Wiley, 2019), over 500 papers and 20 patents.

He is a Fellow of IEEE, and the Editor-in-Chief for *IEEE Antennas and Wireless Propagation Letters*. He was a Distinguished Lecturer of IEEE Antennas and Propagation Society (2014-2016), and an Associate Editor for several international Journals (*IEEE TAP*; *Radio Science*; *Electronics Letters*; *IET Circuits, Devices and Systems*, etc). He served as the Lead Guest Editor of *Proceedings of the IEEE* for a Special Issue on “Small Satellites” (2018), and the Lead Guest Editor of *IEEE Trans on Antennas and Propagation* for Special Issues on "Low-Cost Wide-Angle Beam-Scanning Antennas” (2022) and "Antennas for Satellite Communication" (2015), and a Guest Editor of *IET Circuits, Devices & Systems* for a Special Issue in “Photonic and RF Communications Systems” (2014). He was the UK’s Representative in European Association on Antennas and Propagation (EurAAP) during 2021-2022, General Chair or General Co-Chair of international conferences (LAPC 2013, UCMMT 2021) and was an Invited/Keynote Speaker at many conferences. He is TPC Chair of ISAPE 2024.